\documentclass[prb,aps,twocolumn,amsmath,amssymb,superscriptaddress]{revtex4}
\usepackage{float}
\usepackage{amsmath}
\usepackage{amssymb}
\usepackage{amsfonts}
\usepackage{euscript}
\usepackage{enumerate}
\usepackage{hhline}
\usepackage{pslatex}
\usepackage{tabularx}
\usepackage{color}

\usepackage{graphicx}
\usepackage{dcolumn}
\usepackage{bm}
\usepackage[sort&compress]{natbib}

\makeatletter
\renewcommand{\@biblabel}[1]{#1. }
\renewcommand{\@dotsep}{500}
\renewcommand{\@pnumwidth}{0em}
\renewcommand{\l@figure}[2]{
\@dottedtocline{1}{1.5em}{2em}{Figure #1}{}\vspace{15pt}}

\begin{document}

\title{Nanoscale optical positioning of single quantum dots for bright and pure single-photon emission}

\author{Luca Sapienza}\email{l.sapienza@soton.ac.uk}
\affiliation{Center for Nanoscale Science and Technology, National
Institute of Standards and Technology, Gaithersburg, MD 20899,
USA}\affiliation{Maryland NanoCenter, University of Maryland,
College Park, MD 20742, USA}
\affiliation{School of Physics and Astronomy, University of Southampton, Southampton SO17 1BJ, United Kingdom}
\author{Marcelo Davan\c co}
\affiliation{Center for Nanoscale Science and Technology, National
Institute of Standards and Technology, Gaithersburg, MD 20899,
USA}
\author{Antonio Badolato}
\affiliation{Department of Physics and Astronomy, University of
Rochester, Rochester, NY 14627, USA}
\author{Kartik Srinivasan} \email{kartik.srinivasan@nist.gov}
\affiliation{Center for Nanoscale Science and Technology, National
Institute of Standards and Technology, Gaithersburg, MD 20899, USA}

\date{\today}

\begin{abstract}
\textbf{Self-assembled, epitaxially-grown InAs/GaAs quantum dots are promising semiconductor quantum emitters that can be integrated on a chip for a variety of photonic quantum information science applications.  However, self-assembled growth results in an essentially random in-plane spatial distribution of quantum dots, presenting a challenge in creating devices that exploit the strong interaction of single quantum dots with highly confined optical modes.  Here, we present a photoluminescence imaging approach for locating single quantum dots with respect to alignment features with an average position uncertainty $<30$~nm ($<10$~nm when using a solid immersion lens), which represents an enabling technology for the creation of optimized single quantum dot devices.  To that end, we create quantum dot single-photon sources, based on a circular Bragg grating geometry, that simultaneously exhibit high collection efficiency (48~$\%~\pm$~5~$\%$ into a 0.4 numerical aperture lens, close to the theoretically predicted value of 50~$\%$), low multiphoton probability ($g^{(2)}(0)<$1~$\%$), and a significant Purcell enhancement factor ($\approx$~3)}.
\end{abstract}

 \maketitle

\maketitle

Single InAs/GaAs quantum dots are one of the more promising solid-state quantum emitters for applications such as quantum light generation and single-photon level nonlinear optics~\cite{ref:Michler_book_2009}.  Critical to many such applications is the incorporation of the quantum dot within an engineered photonic environment so that the quantum dot interacts with only specific optical modes.  A variety of geometries such as photonic crystal devices and whispering gallery mode resonators have been employed to achieve such behavior for bright single-photon sources and strongly coupled quantum dot-cavity systems~\cite{ref:Santori_Book}. The optical field in many such geometries varies significantly over distances of $\approx$~100~nm, setting a scale for how accurately the quantum dot position should be controlled within the device for optimal interaction.  While site-controlled growth of quantum dots presents one attractive option~\cite{ref:Juska_site_controlled_QD_entangled}, the properties of such quantum dots (in terms of homogeneous linewidth, for example) have not yet matched those of quantum dots grown by strain-mediated self-assembly (Stranski-Krastanow growth)~\cite{ref:Huggenberger_Forchel_site_controlled}.  However, the in-plane location, polarization, and emission wavelength of such self-assembled quantum dots are not accurately controlled in a deterministic fashion, and thus techniques are required to determine these properties prior to device fabrication, in order to create optimally performing systems. Several techniques for location of self-assembled InAs/GaAs quantum dots prior to device fabrication have been reported, including atomic force microscopy (AFM)~\cite{ref:Hennessy3}, scanning confocal photoluminescence microscopy~\cite{ref:Thon_APL_09} (including in-situ, cryogenic photolithography~\cite{ref:Lee_Taylor_cryo_litho,ref:Dousse_Senellart_QD_in_situ_litho}), photoluminescence imaging~\cite{ref:Kojima_Noda_positioning}, and scanning cathodoluminescence~\cite{ref:Gschrey_Reitzenstein_CL_positioning}. Of these approaches, photoluminescence imaging is particularly attractive given its potential to combine high throughput sub-50~nm positioning accuracy, spectral information, and compatibility with high-resolution electron-beam lithography that is typically used to pattern small features such as those used in photonic crystals.  Localization of single molecules to 10~nm scale accuracy by imaging their fluorescence onto a sensitive camera has proven to be a powerful technique in the biological sciences~\cite{ref:Thompson_Webb_localization}.

Here, we present a two-color photoluminescence imaging technique to determine the position of single quantum dots with respect to fiducial alignment marks, with an average position uncertainty $<30$~nm obtained for an image acquisition time of 120~s (the average position uncertainty is reduced to $<10$~nm when using a solid immersion lens).  This wide-field technique is combined with confocal measurements within the same experimental setup to determine emission wavelength and polarization.  We use this information to fabricate and demonstrate quantum dot single-photon sources in a circular Bragg grating geometry that simultaneously exhibit high collection efficiency (48~$\%~\pm$~5~$\%$ into a lens with numerical aperture of 0.4), low multiphoton probability at this collection efficiency ($g^{(2)}(0)<$1~$\%$), and a significant Purcell enhancement factor ($\approx$~3).  Our results constitute an important step forward for both the general creation of nanophotonic devices using positioned quantum dots, and the specific performance of quantum dot single-photon sources.

\noindent \textbf{\large{Results}}

\noindent \textbf{Quantum dot location via photoluminescence imaging}

\begin{figure*}
\begin{center}
\includegraphics[width=0.75\linewidth]{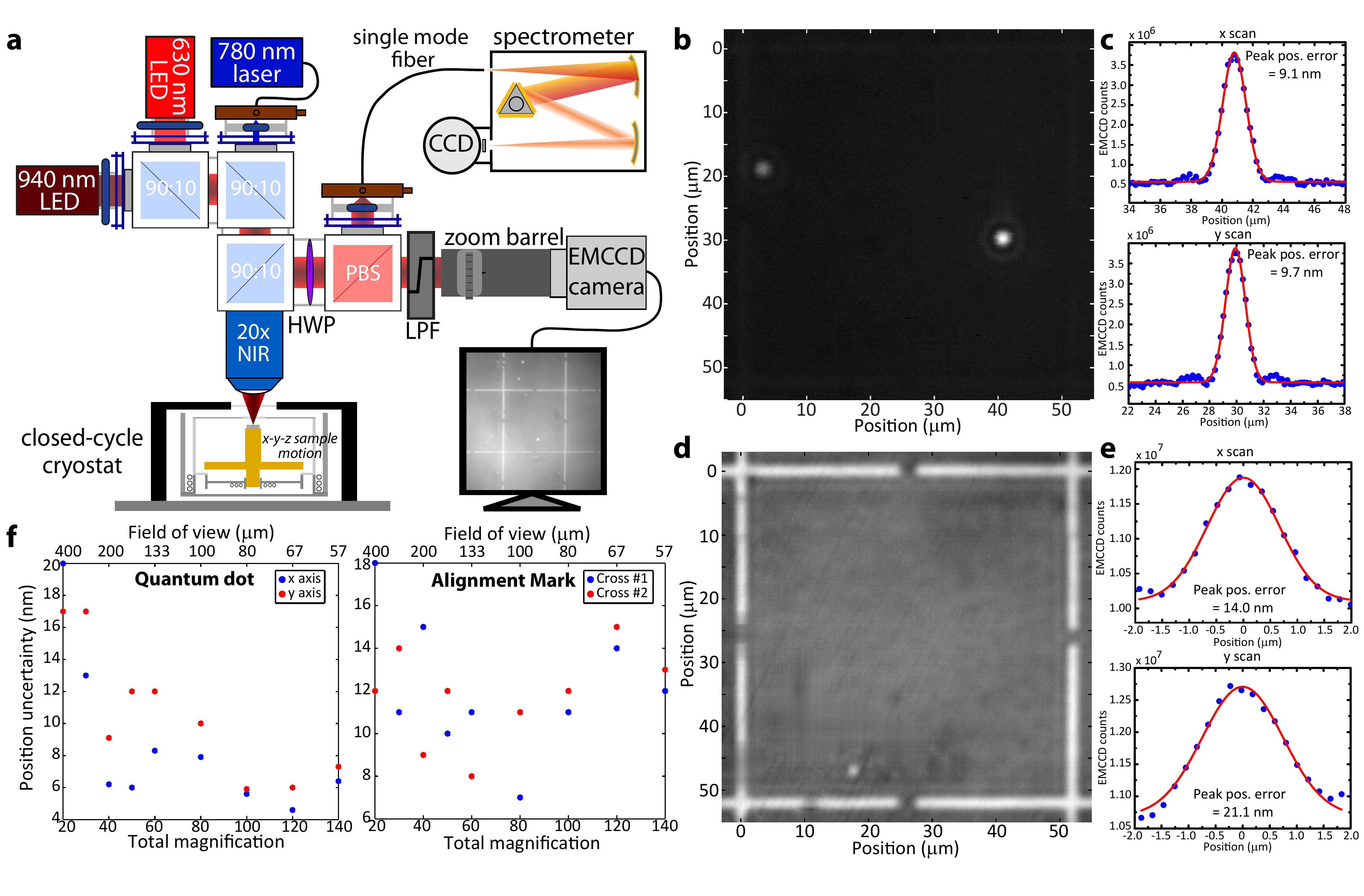}
\caption{\textbf{Optically locating single quantum dots.} (a) Schematic of the photoluminescence imaging setup. An infrared light emitting diode (LED, emission centered at 940~nm) is used for illumination of the sample while either a 630~nm red LED or a 780~nm laser is used for excitation of the quantum dots, depending on whether excitation over a broad area (LED) or of individual quantum dots (laser) is required. Samples are placed within a cryostat on an x-y-z positioner. Imaging is done by directing the emitted and reflected light into an Electron Multiplied CCD (EMCCD) camera, while spectroscopy is performed by collecting emission into a single-mode fiber and sending it to a grating spectrometer. (b) Example photoluminescence image from single quantum dots measured under red LED illumination only.  A 900~nm long pass filter (LPF) is inserted into the collection path when measuring the quantum dot emission. (c) Two orthogonal line cuts (horizontal = x-axis, vertical = y-axis) of the photoluminescence image, showing the profiles of the quantum dot emission (symbols) and their Gaussian fits (lines). (d) Example image of the reflected light from the metallic alignment marks under red LED illumination only.  (e) Two orthogonal line cuts (horizontal = x-axis, vertical = y-axis) of the image in (d), showing the profiles of the reflected light from the metallic alignment marks (symbols) and their Gaussian fits (lines). (f) Peak position uncertainties measured from the Gaussian fits of linecuts of the EMCCD images, plotted as a function of magnification and field of view for the quantum dot and metallic alignment marks.  The uncertainties represent one standard deviation values determined by a nonlinear least squares fit of the data.}
\label{fig:Fig1}
\end{center}
\end{figure*}

Prior to sample interrogation, an array of metal alignment marks is fabricated on quantum-dot-containing material through a standard lift-off process (see Methods). The samples are then placed on a stack of piezo-electric stages to allow motion along three orthogonal axes (x,y,z) within a closed-cycle cryostat that reaches temperatures as low as 6~K. The simplest photoluminescence imaging configuration we use is a subset of Fig.~\ref{fig:Fig1}(a), and starts with excitation by a 630~nm LED, which is sent through a 90/10 (reflection/transmission percentage) beamsplitter and through a 20x infinity-corrected objective (0.4 numerical aperture) to produce an $\approx$~200~$\mu$m diameter spot on the sample.  Reflected light and fluorescence from the sample goes back through the 90/10 beamsplitter and is imaged onto an Electron Multiplied Charged Couple Device (EMCCD) using a variable zoom system.  When imaging the fluorescence from the quantum dots, the 630~nm LED power is set to its maximum power ($\approx$~40~mW, corresponding to an intensity of $\approx$~130~W/cm$^2$), and a 900~nm long-pass filter (LPF) is inserted in front of the EMCCD camera to remove reflected 630~nm light.  Imaging of the alignment marks is done by reducing the LED power to 0.8~mW, turning off the EMCCD gain, and removing the 900~nm LPF.

\begin{figure*}
\begin{center}
\includegraphics[width=0.7\linewidth]{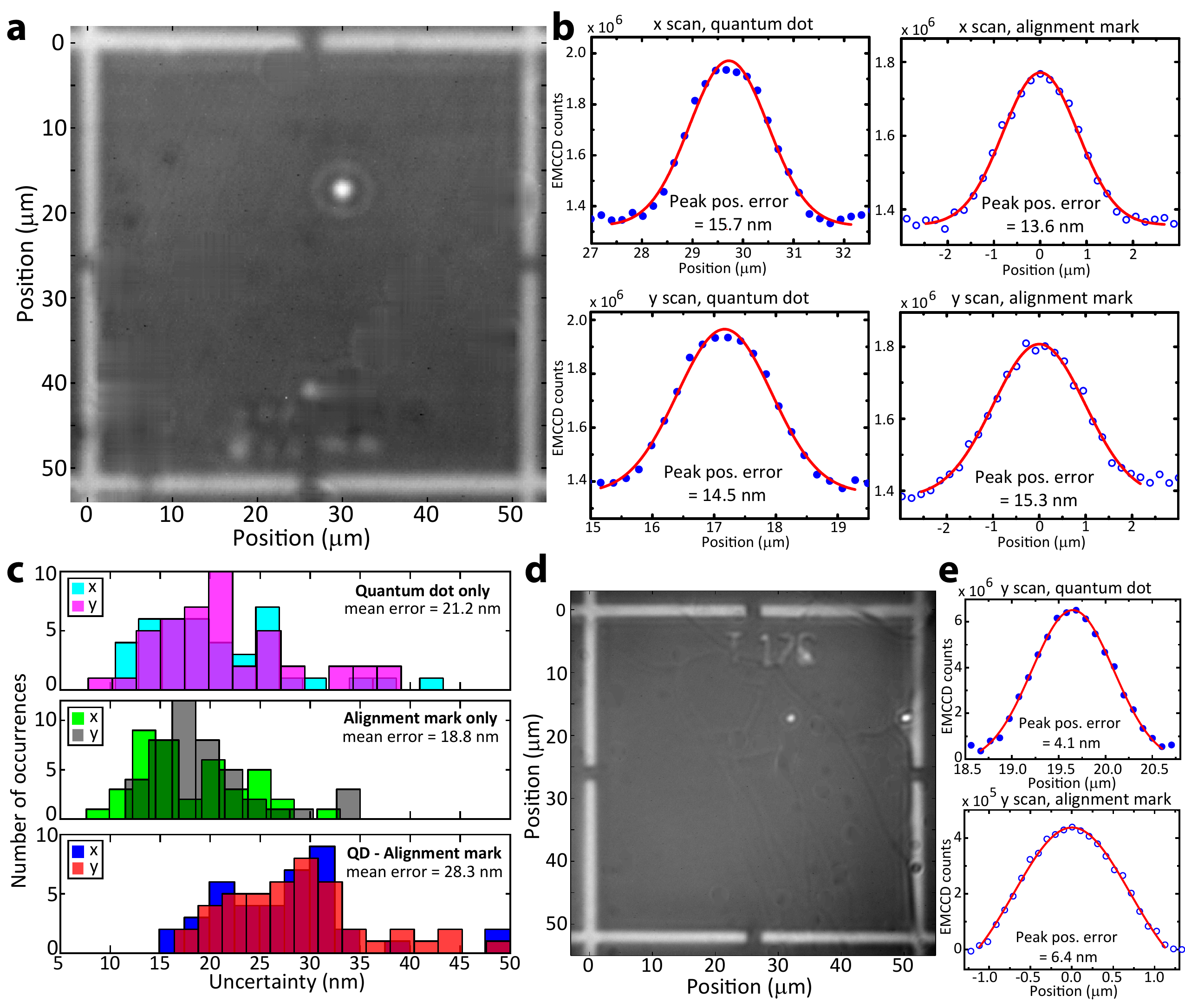}
\caption{\textbf{Performance of the two-color positioning technique.} (a) EMCCD image of the photoluminescence from a single quantum dot and reflected light by the alignment marks (metallic crosses), acquired by illuminating the sample simultaneously with both the red and near-infrared LEDs. (b) Orthogonal line cuts (horizontal=x axis, vertical=y axis) of the photoluminescence image, showing the profiles of the quantum dot emission (solid symbols) and of the image of the alignment marks (open symbols) and their Gaussian fits (solid lines). (c) Histograms of the uncertainties of the quantum dot and alignment mark postions and quantum dot-alignment mark separations, measured from the Gaussian fits of linecuts from 45 images.  The uncertainties represent one standard deviation values determined by a nonlinear least squares fit of the data.  (d)-(e) Photoluminescence imaging through a solid immersion lens. (d) Image of the photoluminescence from single quantum dots and reflected light from the alignment marks (metallic crosses), collected under the 630~nm/940~nm co-illumination scheme. (e) Y-axis line cuts from the photoluminescence image, showing the profiles of the quantum dot emission (solid symbols) and reflected light from the alignment mark (open symbols).  The solid lines are nonlinear least squares fits to Gaussians.}
\label{fig:Fig2}
\end{center}
\end{figure*}

Representative images of the quantum dot photoluminescence and alignment marks are shown in Fig.~\ref{fig:Fig1}(b) and Fig.~\ref{fig:Fig1}(d).  In Fig.~\ref{fig:Fig1}(b), circular bright spots surrounded by Airy rings - a signature of optimally focused collection - are clearly visible and represent the emission from single quantum dots excited within an $\approx$ 56~$\mu$m x 56~$\mu$m field of view. Orthogonal linescans of the bright spots (Fig.~\ref{fig:Fig1}(c)) are fit with Gaussian functions using a nonlinear least squares approach (see Supplementary Note 1), with the extracted peak positions showing one standard deviation uncertainties as low as $\approx$~9~nm.  A similar analysis of orthogonal linescans of the alignment marks (Fig.~\ref{fig:Fig1}(d)-(e)) shows their center positions to be known with an uncertainty that is typically $\approx$~15~nm.  Figure~\ref{fig:Fig1}(f) shows how this uncertainty changes as a function of system magnification (and hence field of view), which is adjusted using the variable zoom system.  We see that the quantum dot uncertainty values show a decreasing trend with higher magnification, and values as low as $\approx$~5~nm are measured.  This can be understood because the increased magnification spreads the quantum dot emission over a larger number of pixels on the EMCCD camera, resulting in a smaller fit uncertainty, provided that the collected fluorescence level produces an adequate per pixel signal-to-noise level. On the other hand, the uncertainty in the alignment mark center position shows no obvious trend with changing magnification.  Ultimately, we have found that the alignment mark uncertainties are limited by the blur induced by the two intermediate fused silica cryostat windows (vacuum and radiation shield, 2~mm and 1~mm thick, respectively) between the objective and sample, which has been confirmed by measurements in ambient conditions with the windows removed.

While the 630~nm LED can thus be used for imaging both the quantum dots and alignment marks, it requires the acquisition of two separate images, with insertion of a filter needed when collecting the quantum dot photoluminescence.  As filter insertion can result in beam shifts that will be manifested as an uncontrolled error in determination of the separation between quantum dot and alignment mark, we implement a modified setup (Fig.~\ref{fig:Fig1}(a)) in which a second, infrared LED at 940~nm is combined with the 630~nm LED when illuminating the sample.  Unlike the 630~nm LED, the 940~nm LED does not excite the quantum dots, but instead serves only to illuminate the alignment marks, with the wavelength chosen to approximately match the expected wavelength of the quantum dot emission.  By adjusting the 940~nm LED power appropriately, both the quantum dots and alignment marks can be observed in a single image with the 900~nm LPF in place.

Figure~\ref{fig:Fig2}(a) shows an image taken when the sample is co-illuminated by both 630~nm and 940~nm LEDs, with the 940~nm power set to be $\approx$~4~$\mu$W, about four orders of magnitude smaller than that of the 630~nm LED power. Orthogonal line scans through the quantum dot and alignment marks under this co-illumination scheme are shown in Fig.~\ref{fig:Fig2}(b). As expected, the uncertainty values determined for quantum dot and alignment mark positions are larger than those obtained when acquiring two separate images (Fig.~\ref{fig:Fig1}(c),(e)), for which the LED power can be optimized independently to maximize the image contrast and minimize each uncertainty. However, we have favoured the co-illumination approach due to its ability to reduce some potential uncertainties, like sample drift, that may occur during schemes requiring multiple images to be acquired.  Ultimately, one might envision time-multiplexing and drift compensation techniques being employed to correct for such factors.

After carrying out a systematic study of the position uncertainties as a function of magnification, integration time, and EMCCD gain, we have found optimized settings for image acquisition (under 40x magnification), in terms of the combined quantum dot and alignment mark uncertainty: an integration time of 120~s, an EMCCD gain of 200, and the aforementioned LED powers. Under these conditions, we have studied the uncertainties in the quantum dot position, alignment mark position, and quantum dot-alignment mark separation for a number of different quantum dots on our sample. Histograms of the measured values are reported in Fig.~\ref{fig:Fig2}(c), and show that the mean uncertainty in the quantum dot-alignment mark separation is $\approx$~28~nm.  Finally, we note that in the present setup, the available 630~nm LED power is below that required to saturate the quantum dot emission (a comparison with the saturation counts obtained under laser excitation shows that it is about half the value required). Higher 630~nm LED power would increase the collected photoluminescence and reduce the uncertainty values that we have reported.  This pre-eminent role of collected photon flux is well-established in the single emitter localization literature~\cite{ref:Thompson_Webb_localization}.  We have confirmed it in our experiments by using a solid-immersion lens~\cite{ref:Zwiller_Bjork_JAP,ref:Serrels_SIL}, which can both increase the LED intensity at the quantum dot and the fraction of quantum dot emission that is collected by the microscope objective. Placing a hemispherical lens with refractive index $n$~=~2 on the surface of the sample yields individual quantum dot and alignment mark position uncertainties of $\approx$~5~nm (Fig.~\ref{fig:Fig2}(d)-(e)), so that the overall uncertainty in locating the quantum dot with respect to the alignment mark is $<10$~nm (more details provided in Supplementary Note 2).  In total, we note that the positioning uncertainties that we obtain are 2$\times$ to 5$\times$ smaller than previously reported~\cite{ref:Dousse_Senellart_QD_in_situ_litho,ref:Thon_APL_09,ref:Kojima_Noda_positioning}, and are obtained with a single image, acquired over a 120~s acquisition time, and spanning an area of the sample greater than 100~$\mu$m~$\times$~100~$\mu$m.

\noindent \textbf{Realization of circular Bragg grating bullseye cavities}

\begin{figure}
\begin{center}
\includegraphics[width=\linewidth]{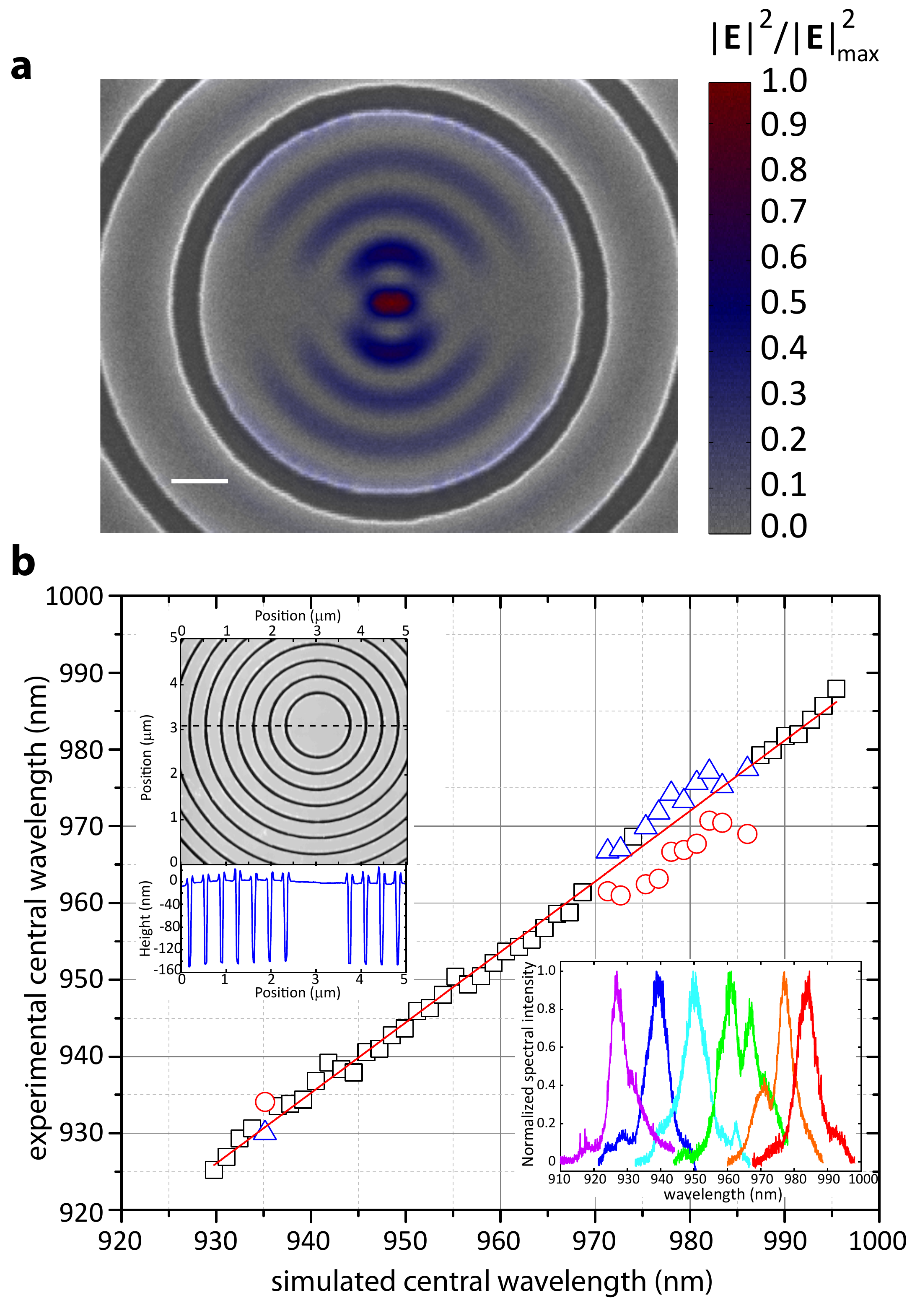}
\caption{\textbf{Circular dielectric gratings tailored to specific quantum dot emitters.} (a) Normalized cavity mode electric field intensity $|E|^2$ superimposed on a scanning electron microscope image of the center of one of the cavities. Scale bar represents 200~nm. (b) Experimental central wavelength of 50 circular grating cavities with varying period and central radius, plotted as a function of the simulated central wavelength. When only one peak is observed in the spectrum, black squares are used to denote the peak wavelength.  When two peaks are observed, red circles and blue triangles are used.  Such two-peak behavior is also seen in simulations depending on the device parameters, and is due to coupling to a second cavity mode. Top inset: Atomic Force Microscope image of a circular grating cavity and a linecut (along the dashed line) showing the etch depth of the trenches. Bottom inset: Examples of photoluminescence spectra of circular grating cavities, measured from a high-quantum dot density region.}
\label{fig:Fig3}
\end{center}
\end{figure}

We now use the optical positioning technique to fabricate nanophotonic structures tailored for the properties of a specific quantum dot and engineered to enhance the collection efficiency of single photons in free space.  First, we obtain information about the quantum dot emission wavelength by spatially selecting one quantum dot and collecting its emission into a single-mode fiber that is coupled into a grating spectrometer (a half waveplate and polarizing beamsplitter are used to switch the collection path between the EMCCD camera and single-mode fiber).  Spatial selection is achieved by exciting individual quantum dots with a 780~nm laser, incorporated into the same micro-photoluminescence setup (Fig.~\ref{fig:Fig1}(a)), and producing a focused spot size of $\approx$~2~$\mu$m on the sample surface. The half waveplate and polarizing beamsplitter also enable determination of the quantum dot polarization.  Having thus obtained emission wavelength to go along with the spatial position obtained from the imaging setup, a properly calibrated fabrication process can enable the creation of nanophotonic structures that are tailored to the specific emitter properties. This allows one to minimise (and potentially avoid altogether) the need for mutual spectral tuning of the emitter with respect to the optical resonance of the cavity, which is a clear limitation of the scalability of these sources.

The specific nanophotonic structure we focus on is a circular Bragg grating `bullseye' geometry, which has been developed as a planar structure in which quantum dot photons are funneled into a near-Gaussian far-field pattern over a moderate spectral bandwidth (few nm) with high efficiency (theoretical efficiency of 50~$\%$ into a 0.4 numerical aperture) and with the potential for Purcell enhancement of the radiative rate~\cite{ref:Davanco_BE,ref:Ates_JSTQE}.  The cavity mode of interest is tightly confined, and optimal performance requires the quantum dot to be within a couple hundred nanometres of the centre of the bullseye structure. This is illustrated in Fig.~\ref{fig:Fig3}(a), which plots the normalized electric field intensity superimposed on a scanning electron microscope image of the center of a fabricated device. An important parameter in the fabrication of these devices is the etch depth of the asymmetric grating, as this determines the fraction of emission in the upwards direction (towards our collection optics) compared to the downwards direction (towards the substrate). Furthermore, given the high refractive index difference between GaAs and air, a change in etch depth of 1~nm results in a shift of the optical resonances of about 1~nm.  We use AFM to determine the GaAs dry etch rate within the grating grooves (Fig.~\ref{fig:Fig3}(b), top inset), and based on this calibration, we fabricate (see Methods) 50 circular gratings whose parameters (pitch and central diameter) have been adjusted so that the cavity resonances cover the 930~nm to 1000~nm range of wavelengths. These samples were fabricated in a region of the wafer with a high density of quantum dots, so that the resulting emission under high power excitation is broad enough to feed the cavity modes. Example spectra collected from different circular grating cavities are shown in the bottom inset of Fig.~\ref{fig:Fig3}(b). These measurements allow us to calibrate the experimental cavity resonances with respect to simulations, as shown in the main panel of Fig.~\ref{fig:Fig3}(b), and tailor the design to match the specific quantum dot emission wavelength.

\noindent \textbf{Optimized quantum dot single-photon source}

\begin{figure*}
\begin{center}
\includegraphics[width=0.8\linewidth]{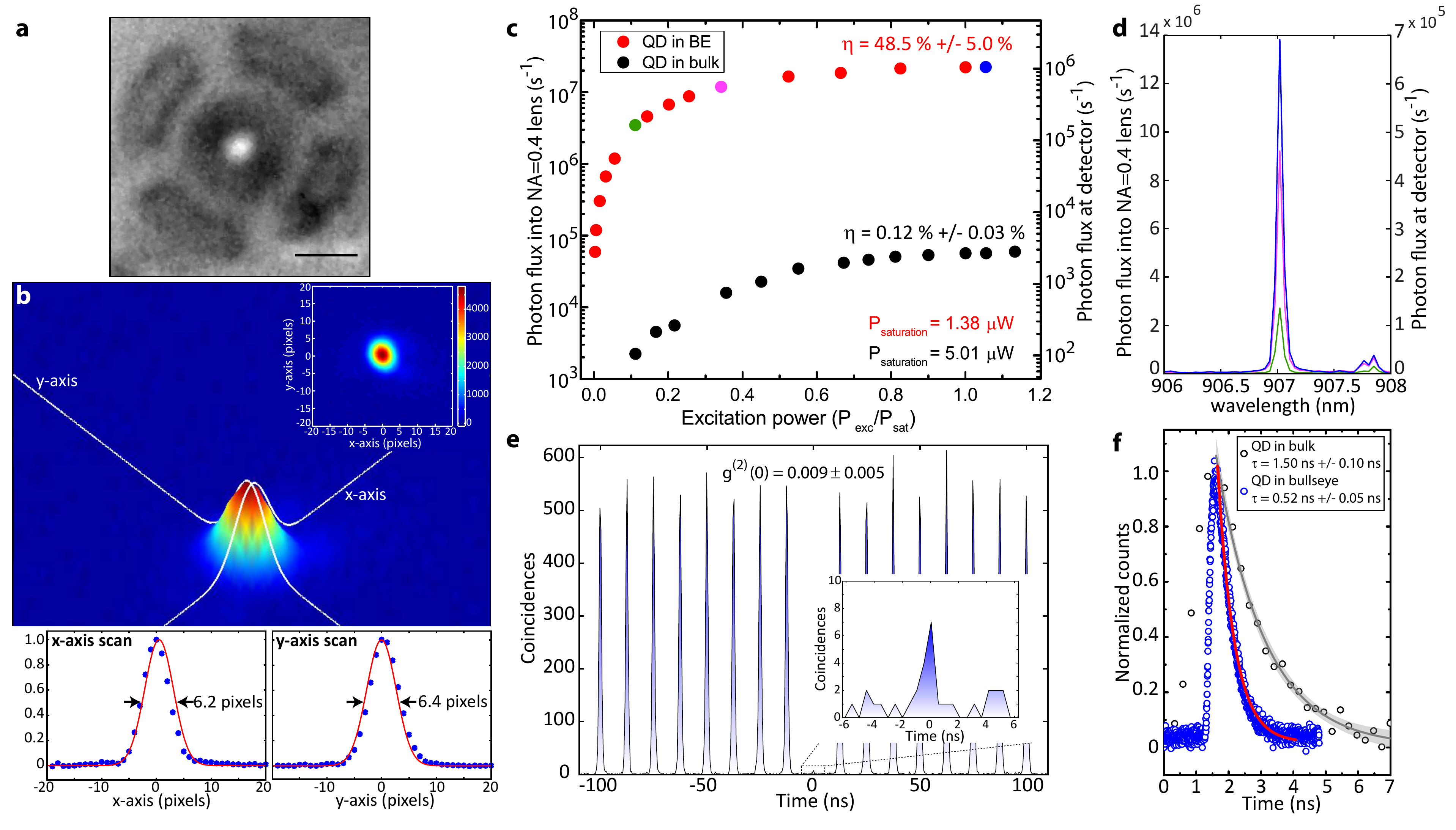}
\caption{\textbf{Single-photon emission from an optimised device.} (a) Image of the photoluminescence from a single quantum dot within the cavity, collected under 630~nm LED illumination. Scale bar represents 5~$\mu$m. (b) Far-field image of the photoluminescence from a quantum dot in a circular grating cavity, along with linecuts from the 2D Gaussian fit to the data along the x- and y-axes, shown as solid white lines.  The upper right inset shows a 2D image plot of the interpolated data, while the bottom curves plot the (uninterpolated) experimental data (symbols) and their Gaussian fits (solid lines). (c) Photon flux into the 0.4 numerical aperture collection objective (left y-axis) and at the detector (right y-axis), plotted as a function of 780~nm excitation power (in saturation units), for a quantum dot (QD) in a circular grating (QD in BE, red symbols) and in unpatterned GaAs (QD in bulk, black symbols). (d) Examples of photoluminescence spectra collected under different excitation power [color coded in panel (c)]. (e) Photon collection coincidence events measured under pulsed 857~nm excitation, using a Hanbury-Brown and Twiss setup. The disappearance of the central peak (zoomed-in plot in the inset) is the signature of pure single-photon emission. The uncertainty value is given by the standard deviation in the area of the peaks away from time zero. See Supplementary Fig.~2 for additional relevant data. (f) Time-resolved photoluminescence measurements collected under pulsed 780~nm excitation, showing the excited state decays (symbols) fitted by single exponential curves (solid lines). The shaded gray areas correspond to the 95~$\%$ confidence intervals in the fit.}
\label{fig:Fig4}
\end{center}
\end{figure*}

Using the quantum dot positions with respect to alignment marks as determined by photoluminescence imaging, emission wavelengths as determined by grating spectrometer measurements, and the aforementioned calibration of the circular grating geometry to match target wavelengths, we fabricate (see Methods) a series of circular grating cavities containing single quantum dots.  Photoluminescence imaging of the devices after fabrication, as shown in Fig.~\ref{fig:Fig4}(a) for a representative device excited by the 630~nm LED, qualitatively indicates that the quantum dot emission originates from the centre of the bullseye structure, as intended.  A measurement of the far-field emission from the device on the EMCCD, as shown in Fig.~\ref{fig:Fig4}(b), shows that it is close to a circular Gaussian function, as confirmed by a nonlinear least squares fit. As the overlap with a perfect circular Gaussian is $\approx$~70~$\%$, this far-field patten is expected to mode match well to a single-mode fiber, an important consideration for long-distance transmission of single photons for quantum information applications.

We now characterise the emission produced by the optically positioned quantum dots within the circular grating cavities, in terms of collection efficiency, single-photon purity, and spontaneous emission rate.  For these measurements, a second cryostat and photoluminescence setup was used, as it provides direct free-space in-coupling to a grating spectrometer that is also used for spectral isolation of the quantum dot excitonic state (Supplementary Fig.~1).  First, we determine the collection efficiency by pumping the devices with a 780~nm wavelength, 50~MHz repetition rate pulsed laser (50~ps pulse width), and varying the laser power until the emission from the quantum dot saturates (Fig.~\ref{fig:Fig4}(c)).  Assuming a quantum dot radiative efficiency of unity, and taking into account the losses within the optical setup (see Supplementary Note~3), we measure a collection efficiency as high as 48.5~$\%$~$\pm$~5.0~$\%$ into a 0.4 numerical aperture objective, where the uncertainty is due to fluctuations in power measurements done to calibrate losses in the optical setup, and represents a one standard deviation value.  This collection efficiency is close to the theoretical value of 50~$\%$ expected for a centrally located quantum dot, and is more than two orders of magnitude larger than the collection efficiency for a quantum dot in unpatterned GaAs, as shown in Fig.~\ref{fig:Fig4}(c). We note that a 80~$\%$ collection efficiency is theoretically expected if a higher numerical aperture optic (e.g., NA~=~0.7) is used.

In previous studies of quantum dots in circular grating cavities~\cite{ref:Davanco_BE,ref:Ates_JSTQE}, where no optical positioning was used, device fabrication in a material containing a higher density of quantum dots was performed, to ensure that some non-negligible fraction (which turned out to be a few percent) of devices would have a quantum dot spectrally and spatially overlapped with the desired cavity mode (see Supplementary Note~4).  In comparison, the optical positioning used here allows us to work with a much lower density of quantum dots ($\lesssim$~1 per 1000 $\mu$m$^2$).  One consequence of this is the comparatively clean emission spectra we observe, even when exciting with pump powers that completely saturate the quantum dot emission (Fig.~\ref{fig:Fig4}(d)).  Such clean spectra might be expected to correspond to clean (low multi-photon probability) single-photon emission, and to test this, the spectrally filtered emission from the bright quantum dot exciton line is measured in a standard Hanbury-Brown and Twiss setup.  Under non-resonant, 780~nm pulsed excitation, we measure $g^{(2)}(0) = 0.15 \pm 0.03$ when the quantum dot emission is saturated (Supplementary Fig.~2(a)).  When the system is excited above quantum dot saturation non-resonantly, we observe emission from the bullseye cavity modes superimposed with the quantum dot emission (data in Supplementary Fig.~2(b), collected under continuous wave 780~nm excitation).  Together, this suggests that quasi-continuum states, originating from the combined single quantum dot - wetting layer system, feed the optical cavity mode~\cite{ref:Winger2009,ref:Chauvin_Fiore,ref:Laucht_Finely_2010} and limit the device's single-photon purity.

We next consider pumping the device on an excited state of the quantum dot, as such excitation (sometimes referred to as quasi-resonant or p-shell pumping) has been shown to reduce $g^{(2)}(0)$~\cite{ref:Santori_NJP}.  Measurement of the quantum dot emission under pulsed 857~nm excitation shows that, at the saturation pump intensity (where the collection efficiency is maximized), the spectrum is nearly identical to that under 780~nm excitation (Supplementary Fig.~2(d)).  Moreover, increased excitation power above saturation (achieved using a 857~nm continuous wave laser) yields far less cavity mode feeding than in the corresponding 780~nm case (Supplementary Fig.~2(c)), suggesting that improved single-photon purity should be observed.  This is confirmed by intensity autocorrelation measurements, which indicate that on-demand single-photon emission with a purity of 99.1$\%$ ($g^{(2)}(0) = 0.009 \pm 0.005$) is achieved at quantum dot saturation. We note that the $g^{(2)}(0)$ levels are determined from raw coincidences, without any background subtraction, and with an uncertainty value given by the standard deviation in the area of the peaks away from time zero.

We also measure the spontaneous decay rate of the quantum dot emission under 780~nm pulsed excitation (measurements at 857~nm have also been performed and yield unchanged results). The spontaneous emission decay of a quantum dot in bulk and a quantum dot in a circular grating cavity are shown in Fig.~\ref{fig:Fig4}(f). The exponential fit of the decay curve allows us to extract a lifetime of $\approx$~520~ps for the quantum dot in the bullseye cavity, corresponding to a Purcell enhancement of the spontaneous emission rate by a factor of $\approx$~3.  A Purcell factor as high as 4 is measured in other devices that have a smaller detuning with respect to the cavity mode (the detuning is 1.6~nm for the device we focus on here). Theoretically, Purcell factors as high as $\approx$~11 are expected~\cite{ref:Davanco_BE,ref:Ates_JSTQE} for quantum dots with perfect spectral and spatial alignment with respect to the cavity mode. Different methods to achieve such precise spectral resonance are currently under consideration; preliminary measurements indicate that in-situ $N_2$ deposition is ill-suited to the circular grating geometry, as the cavity mode degrades before a significant wavelength shift is observed.

Going forward, it would be relevant to determine the location of the optically positioned quantum dots within fabricated devices, in order to understand sources of error within our overall fabrication approach (which combines optical positioning with aligned electron-beam lithography). Supplementary Note~5 presents a detailed discussion on the results of finite-difference time-domain simulations examining the Purcell factor, collection efficiency, and degree of polarization in the collected far-field as a function of dipole position and orientation within the cavity.  Our calculations indicate that the Purcell enhancement, in particular, very sensitively depends on the dipole location, while the collection efficiency is not as sensitive.  For the devices we have focused on in the main text, we find that a simulated offset between 50~nm and 250~nm with respect to the cavity center produces results that are consistent with our measurements.

\noindent \textbf{\large{Discussion}}

There has been much progress in the development of bright quantum dot single-photon sources in recent years, including micropillar~\cite{ref:Strauf_NPhot,ref:Gazzano_Senellart_QD_SPS}, vertical nanowire waveguide~\cite{ref:Claudon,ref:Reimer_Zwiller_nanowire_SPS,ref:Kremer_Gerardot_QD_nanowire_SPS}, fiber-coupled microdisk~\cite{ref:Ates_Srinivasan_Sci_Rep}, and photonic crystal cavity~\cite{ref:Madsen_Lodahl_QD_PC_SPS} geometries.  Many metrics are needed to characterize these sources, and the choice of which ones are of particular importance is largely determined by the intended application. Within the landscape of these sources, the results presented here are unique in terms of simultaneously exhibiting high collection efficiency, nearly perfect single-photon purity at the highest measured collection efficiency, and Purcell enhancement of the spontaneous emission rate.  For example, previous bright, Purcell-enhanced microcavity single-photon sources have shown significant non-zero $g^{(2)}(0)$ values (e.g $\gtrsim0.1$) at their highest collection efficiencies~\cite{ref:Strauf_NPhot,ref:Gazzano_Senellart_QD_SPS,ref:Ates_Srinivasan_Sci_Rep,ref:Madsen_Lodahl_QD_PC_SPS}, while bright nanowire sources show $g^{(2)}(0)\approx0$ but do not exhibit Purcell enhancement~\cite{ref:Claudon,ref:Reimer_Zwiller_nanowire_SPS,ref:Kremer_Gerardot_QD_nanowire_SPS}.  For some applications, the metrics demonstrated thus far should be combined with a high degree of photon indistinguishability~\cite{ref:Gazzano_Senellart_QD_SPS}, which is limited in our work by the coherence time of the quantum dots in this sample ($<300$~ps, as confirmed by measurements with a scanning Fabry Perot interferometer; other emitters on the same wafer show coherence times as long as 500~ps). Future work will focus on resonant excitation~\cite{ref:Muller,ref:Ates_PRL09,ref:He_indistinguishable_SPS} to improve the coherence time and fine control of the cavity-quantum dot detuning to achieve shorter radiative lifetimes~\cite{ref:Santori2,ref:Varoutsis_PRB05}. Together, these advances may provide a route to a source that simultaneously provides bright, pure, and indistinguishable single-photons.

In conclusion, we have developed a photoluminescence imaging technique that enables the location of single quantum dots with respect to alignment markers with an average position uncertainty $<$~30~nm and reaching values as low as $<10$~nm.  We have combined this technique with systematic calibration of our fabrication process to create single-photon sources based on a circular Bragg grating geometry that simultaneously exhibit high brightness, purity, and Purcell enhancement of the spontaneous emission rate. More generally, this technique is an important step forward in the ability to create functional single quantum dot nanodevices, including quantum light sources, strongly-coupled quantum dot - microcavity systems for achieving single photon nonlinearities~\cite{ref:Fushman_Vuckovic_Science_phase_shift, ref:Reinhard_Imamoglu_photon_blocakde,ref:Kim_Waks_QD_logic_gate} coupled quantum dot - nanomechanical structures~\cite{ref:Wilson_Rae_Imamoglu_QD_mechanics_theory,ref:Metcalfe_Lawall_QD_SAW,ref:Yeo_strain_QD}, and integrated systems involving multiple quantum dot nodes.

\vspace{3mm} \noindent \large{\textbf{Methods}}

\noindent \small{\textbf{Circular Bragg grating cavity fabrication}}

Devices are fabricated in a wafer grown by molecular beam epitaxy, consisting of a single layer of InAs quantum dots (QDs) embedded in a 190\,nm thick layer of GaAs, which in turn is grown on top of a 1\,$\mu$m thick layer of Al$_x$Ga$_{1-x}$As with an average $x$\,=\,0.65.  The s-shell peak of the QD ensemble is located near 940~nm, and a gradient in the QD density is grown along one axis of the wafer. Low-temperature photoluminescence imaging of portions of the wafer is performed prior to any device definition to determine the appropriate location on the wafer (in terms of QD density) to fabricate devices.

Alignment marks are fabricated using positive tone electron-beam lithography and a lift-off process.  Polymethyl methacrylate (PMMA) with a molecular weight of 495,000 is spin coated onto the sample, and 2~$\mu$m wide, 50~$\mu$m long crosses are patterned in the resist using a 100 keV electron-beam lithography tool.  After exposure, the resist is developed in a 1:3 (by volume) solution of methyl isobutyl ketone (MIBK) and isopropanol, and 20~nm of Cr and 100~nm of Au are deposited on the sample using an electron-beam evaporator. Microposit remover 1165 is used for lift-off, with gentle ultra-sonication applied if necessary.

After location of quantum dots with respect to the alignment marks through photoluminescence imaging, circular Bragg grating `bullseye'  microcavities are fabricated as follows.  First, the sample is spin-coated with a positive tone electron-beam resist (ZEP 520A), and aligned electron-beam lithography with a 100~keV tool and four mark detection is performed.  Next, the pattern is transferred into the GaAs layer using an Ar-Cl$_2$ inductively-coupled plasma reactive ion etch.  After removal of the electron beam resist, the sample is undercut in hydrofluoric acid.

Atomic force microscopy (AFM) was used in the calibration of the etch rate, with the samples scanned in tapping mode using a commercial, etched silicon probe whose backside is coated with Al.  The AFM probe cantilever has a vendor-specified spring constant of 42~N/m, frequency of 300~kHz, and probe tip radius and height of 8~nm and 10~$\mu$m, respectively.


\noindent \textbf{Acknowledgements} L.S. acknowledges
support under the Cooperative Research Agreement between the
University of Maryland and NIST-CNST, Award 70NANB10H193. The
authors thank Serkan Ates and Krishna Coimbatore Balram for useful
discussions and early contributions to this work. They also thank Christopher
Long and Santiago Solares for helpful advice regarding atomic force microscopy.

\noindent \textbf{Author Contributions} L.S. and K.S. fabricated the
devices, performed the measurements and analyzed the experimental data, and wrote the manuscript.
A.B. grew the quantum dot material, M.D. and L.S. performed the electromagnetic
simulations, and K.S. supervised the project.
\\
\noindent The identification of any commercial product or trade name is used to foster understanding. Such identification does not imply recommendation or endorsement or by the National Institute of Standards and Technology, nor does it imply that the materials or equipment identified are necessarily the best available for the purpose.

\noindent \textbf{Additional Information} Correspondence and
requests for materials should be addressed to L.S. and K.S.

\noindent \textbf{Competing financial interests} The authors declare no competing financial interests.

\newpage
\onecolumngrid \bigskip
\appendix
\setcounter{figure}{0}
\setcounter{equation}{0}
\makeatletter
\renewcommand{\theequation}{S\@arabic\c@equation}

\setcounter{figure}{0}
\makeatletter
\renewcommand{\thefigure}{S\@arabic\c@figure}

\begin{center} {{\bf \large SUPPLEMENTARY
INFORMATION}}\end{center}

\noindent \textbf{Supplementary Note 1: Quantum dot positioning setup and measurements}
\vspace{0.1in}

In this note, we provide additional details on the quantum dot positioning setup shown schematically in Fig.~1 of the main text.

The samples are housed within a cryogen-free cryostat with a base temperature as low as 6~K. Sample motion is achieved using a three-axis cryogenic piezo-positioning stage system.  A confocal micro-photoluminescence geometry is utilized, in which a microscope objective (20x magnification and 0.4 numerical aperture) both focuses excitation light on the sample and collects light emitted and reflected by the sample.  As described in the main text, photoluminescence imaging is done with co-illumination by 630~nm and 940~nm LEDs, where the former is used to excite the quantum dots and the latter is used to image alignment marks.  Excitation of single quantum dots for spectroscopy is performed by focusing a 780~nm laser on the sample.

A 90/10 (reflection/transmission percentage) beamsplitter followed by a 900~nm long-pass filter is used to send the light emitted and reflected by the sample towards the imaging and spectroscopic characterisation paths.  Selection between the two paths is accomplished with a half waveplate and polarizing beasmplitter.  For photoluminescence imaging, the collected light is coupled into a variable zoom system and Electron Multiplied Charged Couple Device (EMCCD), while for spectroscopy, it is coupled to a single mode fiber whose output is sent to a grating spectrometer equipped with a silicon Charged Coupled Device (CCD).

In the photoluminescence imaging measurements, the 900~nm longpass filter serves to reject 630~nm excitation light, while allowing both the quantum dot emission and reflected 940~nm LED light to pass. A total system magnification of 40x (20x from the objective, and 2x from the zoom barrel) is used, corresponding to a field of view of $\approx$ 200~$\mu$m x 200~$\mu$m. EMCCD images are acquired with an integration time of 120~s and gain of 200. While the 630~nm LED power is always set at its maximum ($\approx$~40~mW, corresponding to an intensity of $\approx$~130~W/cm$^2$) to generate as much fluorescence from the quantum dot as possible, the 940~nm LED power is set to achieve a reflected signal from the alignment marks that is approximately equal to the intensity of the quantum dot emission.  This choice of 940~nm LED power is a tradeoff between the improved alignment mark position uncertainty produced at higher powers, and the degraded quantum dot position uncertainty that results if the reflected 940~nm LED signal swamps the quantum dot emission. A typical 940~nm LED power is $\approx$~4~$\mu$W.

The linecuts of the images taken by the EMCCD camera are analyzed using a commercial software and fitted by Gaussian functions to determine the location of the quantum dot and centers of the alignment marks. The fit is optimized using a Levenberg Marquardt iteration algorithm. The central position of the Gaussian function and its error are then translated from a pixel value on the camera to a distance on the sample by using a calibration obtained by imaging, under the same magnification conditions, a microscope calibration target presenting etched features with known separations.

\vspace{0.1in}
\noindent \textbf{Supplementary Note 2: Solid immersion lenses for reduced positioning uncertainties}
\vspace{0.1in}

Solid immersion lenses have been used to increase the collection of light emitted by semiconductor quantum dots by increasing the effective numerical aperture of the collection optics~\cite{ref:Serrels_SIL_SI}. Moreover, because the solid immersion lens reduces the focused excitation spot size, it also increases the excitation intensity at the sample.  This can lead to an increased photon flux from the quantum dot, because (as was noted in the main text), without the solid immersion lens, the maximum 630~nm LED excitation intensity at the sample is not enough to saturate the quantum dot emission. Taken together, the increased emission signal from the quantum dot should lead to a lower uncertainty in its position.  We also expect that the solid immersion lens can improve the alignment mark uncertainty, since the amount of 940~nm LED power used to image the mark will be increased (see discussion above) to match the increased quantum dot emission level.

We test the above experimentally using a 2~mm diameter, high refractive index ($n~\approx~2$) half-ball lens placed directly on the sample surface, with a thin layer of cryogenic grease applied between the sample and lens, obtaining the results shown in Fig.~2(d)-(e) from the main text (the x-axis scans are similar). We measure uncertainties in the quantum dot-marker distance as low as 7.6~nm, a reduction of about a factor 4 compared to the average error measured without the lens (and a factor of 2 compared to the best error measured without the lens).

However, there are some considerations to take into account other than the reduced positioning error possible with a solid immersion lens.  First, the solid immersion lenses are generally 1 mm or 2 mm in diameter. Therefore, when using them for imaging the quantum dot emission and alignment marks, the area of the sample that can be probed within a single measurement session is highly reduced, unless multiple lenses are used.  Second, given that the lenses are hemispherical, care must be taken in optimally focusing the imaging and excitation light on the apex of the solid immersion lens, in order to avoid distortions of the image that would affect the inferred distance between the alignment mark and the quantum dot.

\vspace{0.1in}
\noindent \textbf{Supplementary Note 3: Quantum dot single-photon source characterization}
\vspace{0.1in}

A schematic of the experimental setup used to evaluate the collection efficiency and single-photon purity of the quantum dot emission is shown in Supplementary Fig.~\ref{fig:Janis_setup}, and is similar to that used in previous work~\cite{ref:Ates_JSTQE_SI}. The sample is mounted on the cold finger of a liquid helium flow cryostat that sits on a two-axis nano-positioning stage. Spectral properties of the quantum dot emission are investigated via low-temperature micro-photoluminescence, where a 20x microscope objective (numerical aperture of 0.4) is used for both the illumination of the sample and the collection of the emission. Four different excitation sources are available for use.  The first is a continuous wave 780~nm diode laser for basic spectroscopy.  The second is a continuous wave Ti:sapphire laser, tunable beteween 780~nm and 1000~nm, that can be used to excite the quantum dot on its different excited state transitions. The third is a gain-switched, 780~nm pulsed laser diode (50~ps pulse width; 50~MHz repetition rate) for photon counting, lifetime, and correlation measurements.  The final source is a 820 nm to 950 nm tunable fiber laser ($<10$~ps pulse width; 80~MHz repetition rate) used for counting, lifetime, and correlation measurements under excitation of a quantum dot's excited state.

\begin{figure}[h]
\begin{center}
\includegraphics[width=0.6\linewidth]{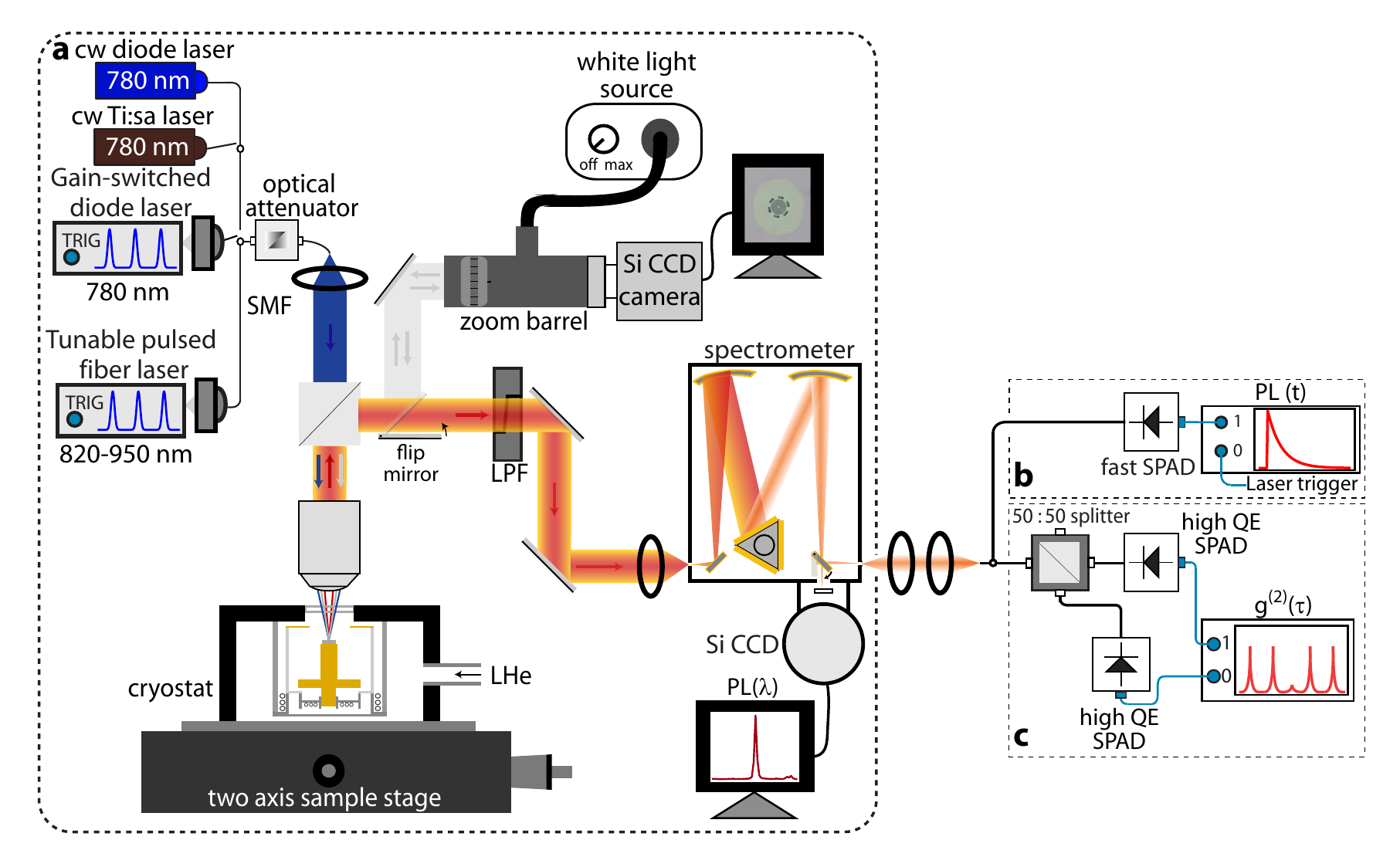}
\caption{Schematic of the experimental setup: (a) confocal micro-photoluminescence; (b) time-resolved photoluminescence; (c) Hanbury Brown and Twiss photon correlation setup.  SMF: single-mode fiber, LPF: long-pass filter, CCD: charge-coupled device, PL: photoluminescence, SPAD: single-photon avalanche diode, QE: quantum efficiency.}
\label{fig:Janis_setup}
\end{center}
\end{figure}

The collected signal is directed to a spectrometer either to record an emission spectrum with a Si CCD camera, or to filter a single emission line for further investigation (Supplementary Fig.~\ref{fig:Janis_setup}(a)). The spectrally filtered emission line is coupled into a single mode optical fiber to enable measurements using fiber-coupled single-photon avalanche diodes (SPADs).  Single quantum dot fluorescence decay dynamics are measured through time-correlated single-photon counting, which relies on measuring the time delay between an excitation pulse and detection of an emitted photon by a SPAD (Supplementary Fig.~\ref{fig:Janis_setup}(b)).  We use a thin Si SPAD whose timing jitter is $<50$~ps to enable measurement of fast quantum dot decay dynamics.  For the second-order correlation function $g^{(2)}(\tau)$ measurements, the spectrally filtered emission is directed to a Hanbury-Brown and Twiss interferometer that consists of a fiber-coupled, 50/50 non-polarizing beam-splitter and two fiber-coupled single-photon avalanche diodes (SPADs), as shown in Supplementary Fig.~\ref{fig:Janis_setup}(c). These SPADs have a timing jitter of $\approx$~700~ps, and their outputs are connected to a time-correlated single-photon counting board.  A time bin width of 512~ps is chosen for the $g^{(2)}(\tau)$ measurements.

Calibration of the quantum dot single-photon source collection efficiency into the 20x (0.4 numerical aperture) objective proceeds as follows.  First, the transmission of the optical path from the QD source to the detector is determined.  The emitted light escapes the cryostat by traveling through two fused silica windows (total transmission $\approx$~87~$\%$), it is then collected by a microscope objective (transmission of $\approx$~70~$\%$), goes through a 90/10 beamsplitter (transmission of $\approx$~89~$\%$), reflects off four dielectric mirrors and travels through a polarizer (total transmission of $\approx$~78~$\%$) before being focused through the slit of the grating spectrometer. The total transmission of the optical path up to the spectrometer is 42~$\%$ $\pm$~4~$\%$, where the uncertainty is based on the spread of transmission values measured for the optical components, and represents a one standard deviation value.

Next, a known laser power (22.6 $\mu$W) is sent into the spectrometer after being attenuated by an independently measured attenuation (64.21~dB) using a variable attenuator, and the detected counts on the Si CCD coupled to the spectrometer are recorded.  The counts measured from a quantum dot, excited with a 50~MHz repetition rate source, are then recorded and compared to the laser counts (taking into account the transmission of the optical path) in order to extract the emitter's single-photon collection efficiency.

\begin{figure}[h]
\begin{center}
\includegraphics[width=\linewidth]{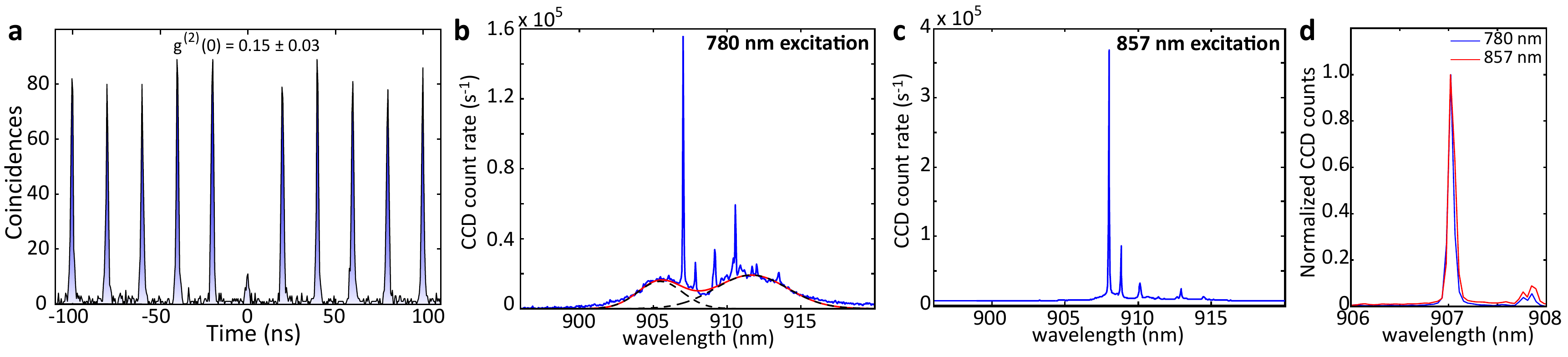}
\caption{Supplementary data for the single photon source characterization of Figure~4 from the main text.  (a) $g^{(2)}(\tau)$ under 780~nm pulsed excitation, with the quantum dot emission saturated. (b) Spectrum under intense 780~nm continuous wave excitation (far above quantum dot saturation), for which the bullseye cavity modes are visible.  The quantum dot line of interest is +1.6~nm detuned with respect to the shorter wavelength cavity mode.  The red solid line is a fit of the data to the sum of two Gaussians, which is used to determine the center wavelengths of the two cavity modes.  The two Gaussians making up the sum are shown as black dashed lines. (c) Spectrum under intense 857~nm continuous wave excitation (far above quantum dot saturation), on resonance with a quantum dot excited state. Compared to 780~nm excitation far above saturation, significantly reduced cavity mode emission is observed. (d) Comparison of the quantum dot spectra under saturation conditions (the conditions under which the $g^{(2)}(\tau)$ data in part (a) and Fig.~4(e) were taken) for both 780~nm and 857~nm pulsed excitation.}
\label{fig:positioned_QD_SPS_SI_data}
\end{center}
\end{figure}

Figure~4 and the accompanying discussion in the main text present data characterizing single-photon source performance for an optically positioned quantum dot within a circular grating `bullseye' cavity.  Here, we present supplementary data referred to in the main text discussion.  Supplementary Fig.~\ref{fig:positioned_QD_SPS_SI_data}(a) shows an intensity autocorrelation measurement ($g^{(2)}(\tau)$) under pulsed excitation at 780~nm when the quantum dot emission is saturated. Despite what appears to be a relatively clean emission spectrum (in terms of an absence of spectral features other than the quantum dot emission) in Fig.~4(d) of the main text, the relatively significant multi-photon component measured ($g^{(2)}(0) = 0.15 \pm 0.03$) indicates the presence of emission that is spectrally resonant with the emission from the quantum dot excitonic line. Quasi-continuum states, generated by hybridization of states of the single quantum dot with those of the wetting layer~\cite{ref:Winger2009_SI,ref:Chauvin_Fiore_SI,ref:Laucht_Finely_2010_SI}, are thought to be a potential source of such multi-photon emission, particularly in Purcell-enhanced (e.g., microcavity) geometries.  This is consistent with our measurements, as more intense excitation (through a 780~nm continuous wave laser that provides more output power than the pulsed 780~nm laser) yields a spectrum in which the cavity mode emission is clearly visible (Supplementary Fig.~\ref{fig:positioned_QD_SPS_SI_data}(b)).  Given that the cavity mode linewidths are many nanometers wide, while the quantum dot excitonic states are orders of magnitude narrower, a spectrally broad emission source such as a quasi-continuum state is needed to reconcile the presence of the cavity mode within the measured spectrum.

The contribution of such quasi-continuum states should be limited if the system is pumped on an excited state of the quantum dot, which would prevent the generation of high energy carriers that could fill those states.  Using a narrow linewidth ($<1$~MHz) continuous wave Ti:sapphire laser, we have identified 857.0~nm and 876.4~nm as wavelengths that are resonant with quantum dot excited states.  Under intense excitation at these wavelengths, the cavity mode feeding is significantly reduced relative to the 780~nm case, as shown in Supplementary Fig.~\ref{fig:positioned_QD_SPS_SI_data}(c) for 857~nm excitation.  Switching to pulsed 857~nm excitation, we find that the resulting spectrum at saturation of the quantum dot emission is nearly identical to that observed under pulsed 780~nm excitation in the main text, as shown in Supplementary Fig.~\ref{fig:positioned_QD_SPS_SI_data}(d). In contrast, the measured $g^{(2)}(\tau)$ (Fig.~4(e) in the main text) is markedly different, with $g^{(2)}(0) = 0.009 \pm 0.005$.  Overall, these results indicate the importance of excited state pumping to achieving pure single photon emission, even in situations in which there is only one quantum dot that can interact with the cavity mode.

Finally, we note that in the photon antibunching experiments, the grating spectrometer was used as a monochromator to spectrally isolate the quantum dot emission, and had a throughput of $\approx$~11~$\%$.  The output of the monochromator was coupled into single mode fiber and sent into the Hanbury-Brown and Twiss setup as described above, and the detected count rates on each of the two SPADs was $\approx~2{\times}10^4$ counts/s in the measurements from Fig. 4(e) in the main text.  Overall, this detected count rate includes the collection efficiency of quantum dot emission into the NA~=~0.4 lens ($\approx$~48~$\%$), the transmission of the photoluminescence setup ($\approx$~42~$\%$), the throughput of the monochromator ($\approx$~11~$\%$), coupling from the monochromator output into single mode fiber and throughput of the single-mode-fiber-based Hanbury-Brown and Twiss setup ($\approx$~12~$\%$), and the SPAD quantum efficiency ($\approx$~20~$\%$).

\vspace{0.1in}
\noindent \textbf{Supplementary Note 4: Comparison to single-photon sources created without optical positioning}
\vspace{0.1in}

For the purposes of comparison, in this section we present data from quantum dot single-photon sources in which quantum dot positioning was not employed (so that the position of the quantum dot with respect to bullseye cavity center was uncontrolled).  The investigation of these devices was described in detail in Ref.~\onlinecite{ref:Ates_JSTQE_SI}, where spectroscopy, lifetime, and photon correlation measurements were presented.  In Supplementary Fig.~\ref{fig:unpositioned_BE_data}(a), we show an EMCCD image of a subset of the array of cavities investigated in Ref.~\onlinecite{ref:Ates_JSTQE_SI}, where the array has been illuminated by the 630~nm red LED. This EMCCD image reveals two new pieces of information. First, only one of twelve displayed devices shows an emission lobe near the center of the cavity, for which the collection efficiency is expected to be maximized.  For this unpositioned sample, the maximum collection efficiency measured was $\approx$~10~$\%$, and the fraction of devices producing this efficiency was a couple of percent. Next, the quantum dot density in this sample is significantly higher than that studied in the current manuscript.  While the density is still low enough so that only a single quantum dot can spatially and spectrally interact with a mode of the cavity, it is about two orders of magnitude larger than what we use in the positioned quantum dot devices. The background emission caused by these quantum dots, and in particular, their potential for supporting quasi-continuum states with broad emission bandwidths~\cite{ref:Winger2009_SI}, may limit the purity of single-photon emission.  Given that the yield for this sample is only a couple of percent, reducing the quantum dot density without locating the quantum dots prior to fabrication is impractical.

\begin{figure}[h]
\begin{center}
\includegraphics[width=0.65\linewidth]{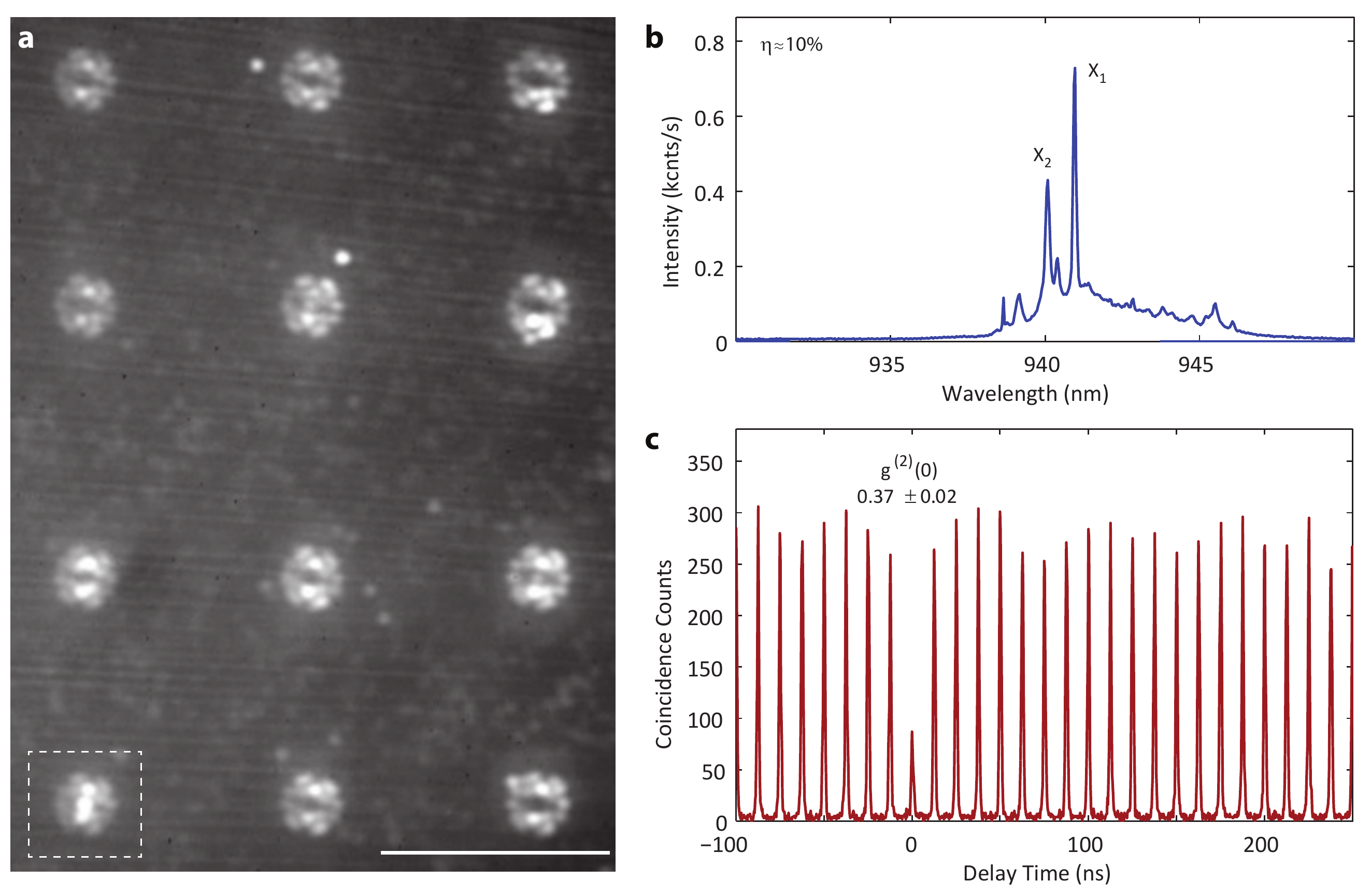}
\caption{Representative data from quantum dot - bullseye cavity devices fabricated without using the quantum dot positioning technique, as in Ref.~\onlinecite{ref:Ates_JSTQE_SI}.  (a) EMCCD image of an array of cavities, illuminated by a 630~nm LED.  Only one of the devices (within the dashed box) shows an emission lobe centered with respect to the cavity. Scale bar represents 50~$\mu$m. (b) Photoluminescence spectrum from a device exhibiting collection efficiency $\approx$~10~$\%$. (c) $g^{(2)}(\tau)$ from the same device.  Note that parts (b) and (c) are re-displayed data from Ref.~\onlinecite{ref:Ates_JSTQE_SI}.}
\label{fig:unpositioned_BE_data}
\end{center}
\end{figure}

Spectroscopy and photon counting measurements from Ref.~\onlinecite{ref:Ates_JSTQE_SI} further address these points.  A typical photoluminescence spectrum under non-resonant pulsed excitation is shown in Supplementary Fig.~\ref{fig:unpositioned_BE_data}(b).  In contrast to the clean spectrum shown in the main text in Fig.~4(d), the spectrum of Supplementary Fig.~\ref{fig:unpositioned_BE_data}(b) shows significant background emission attributed to feeding of the cavity mode by multi-excitonic states of nearby quantum dots.  This emission can be expected to limit the purity of the single-photon source produced by spectrally isolating a single excitonic state, and indeed, the $g^{(2)}(\tau)$ measurement in Supplementary Fig.~\ref{fig:unpositioned_BE_data}(c) shows a significant departure from $g^{(2)}(0)=0$.  While this measurement is of a device with a particularly high $g^{(2)}(0)$ value, in general, unpositioned devices studied in Ref.~\onlinecite{ref:Ates_JSTQE_SI} showed $g^{(2)}(0)\gtrsim15~\%$.  That being said, the discussion from the previous section indicates that even the drastically reduced quantum dot density used in the current work most likely needs to be supplemented by excited state pumping of the quantum dot in order to achieve $g^{(2)}(0)~\approx~0$.

\vspace{0.1in}
\noindent \textbf{Supplementary Note 5: Electromagnetic simulations}
\vspace{0.1in}

As discussed in the main text and in Ref.~\onlinecite{ref:Ates_JSTQE_SI}, the bullseye cavity supports dipole-like resonant modes (shown in Supplementary Figs.~\ref{fig:dipole}(b) and (c)) that are well-suited for the creation of bright single-photon sources - a combination of relatively high Purcell-type radiative enhancement, efficient vertical light extraction from the semiconductor, and near-Gassian far-field for efficient collection into an optical fiber. These modes are strongly localized at the center of the cavity (the central intensity peak has a full-width at half-maximum of $\approx~100$~nm), and a sequence of satellite peaks along the radial direction. Because the electric dipole coupling to a cavity mode is proportional to the squared electric field magnitude {\it at the dipole location}~\cite{ref:Xu_99}, we expect that the Purcell enhancement factor $F_p$, coupling efficiency $\eta$, and emitted polarization state will vary significantly with dipole position. An understanding of these parameters is not just important from a device performance perspective, but also provides information about the actual quantum dot location. We employ full-wave numerical electromagnetic simulations to investigate the sensitivity of the emission properties of our single-photon source to the location of the quantum dot within the bullseye cavity.

\subsection{Purcell Factor and Collection Efficiency}

Following Ref.~\onlinecite{ref:Xu_99}, we use finite-difference time domain (FDTD) simulations to model the system as an electric dipole radiating inside a suspended bullseye cavity. The dipole is allowed to radiate with a short Gaussian pulse time dependence, and the electromagnetic field is allowed to evolve over a long time span. The steady-state electromagnetic field is recorded at all edges of the computational window, so that the total dipole radiated power $P_{rad}$ can be determined. The Purcell factor can then be obtained as $F_p=P_{rad}/P_{hom}$, where $P_{hom}$ is the dipole radiated power in a homogeneous medium~\cite{ref:Xu_99}.  We also record the power $P_z$ emitted upwards in the $+z$ direction, which in a real setting is partially collected with a microscope objective with numerical aperture $NA$. The steady-state field recorded at a parallel plane above the bullseye cavity is used to calculate the emitted far-field, which is then integrated within an angular cone corresponding to a numerical aperture $NA$ to yield the collection efficiency $\eta_{NA}$. Perfectly matched layers are used to simulate free-space above and below the cavity, so that effects related to the substrate are not taken into account. The dipole is assumed to be on the $z=0$ plane (which corresponds to the center of the semiconductor membrane) as defined by the quantum dot growth process, and to have no $z$-components. The latter assumption is appropriate for epitaxially grown InAs dots, given their few nanometer vertical size, negligible compared to the membrane thickness~\cite{ref:Michler_book_2009_SI}.

\begin{figure}[h]
\begin{center}
\includegraphics[scale=1]{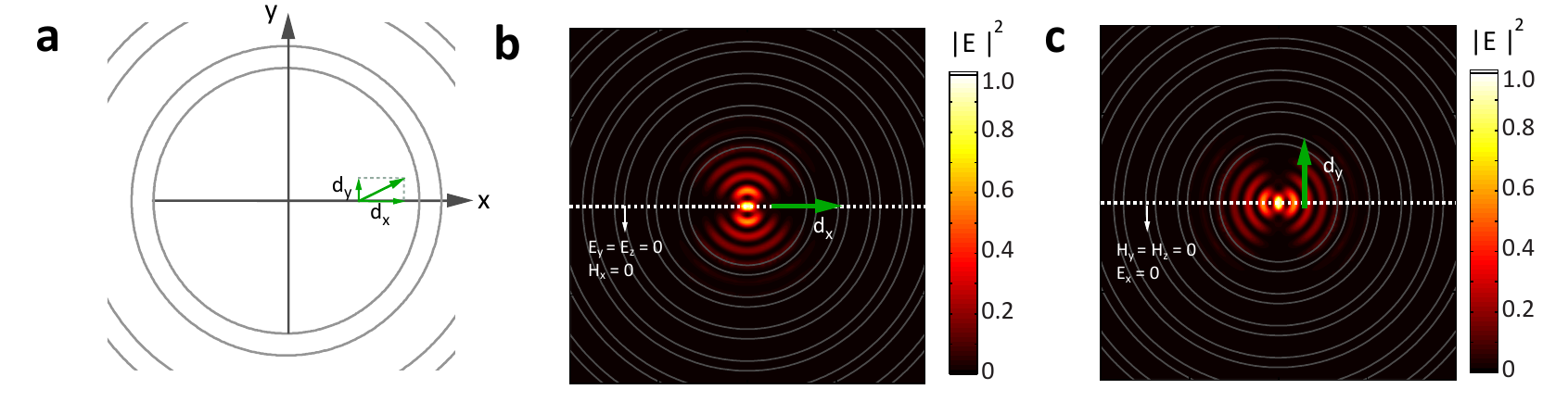}
\caption{(a) Schematic of the simulation geometry, showing a dipole (green arrows) inside a bullseye cavity. Due to the circular symmetry, a dipole located anywhere within the cavity and with any orientation can be represented by a dipole on the $x$-axis with $d_x$ and $d_y$ components equivalent to the radial and azimuthal ones. (b) Electric field amplitude squared profile for a 'h'-type bullseye cavity mode. The $yz$-plane boundary conditions satisfied by the mode are given in the figure. This mode is only excited by $x$-dipoles. (c) Electric field amplitude squared profile for a 'v'-type bullseye cavity mode. The $yz$-plane boundary conditions satisfied by the mode are given in the figure. This mode is only excited by $y$-dipoles.}
\label{fig:dipole}
\end{center}
\end{figure}

In the circular geometry of the cavity, an electric dipole with moment $\mathbf{d}$ with arbitrary orientation placed anywhere in the cavity is equivalent to a dipole located on the $x$-axis with components along the $x$ and $y$ directions, corresponding to the radial and azimuthal components. This is illustrated in Supplementary Fig.~\ref{fig:dipole}(a). The symmetry of the problem allows a description of the electromagnetic fields supported by the cavity in terms of orthogonal, symmetric and anti-symmetric cavity eigenmodes with respect to the $y=0$ plane ('$h$'-modes) and degenerate modes of the 90-degree rotated geometry ('$v$'-modes), as illustrated in Supplementary Figs.~\ref{fig:dipole}(b) and (c). An $x$-dipole will however only excite symmetric $h$-modes and anti-symmetric $v$-modes, and vice-versa is valid for a $y$-dipole; in other words, $\mathbf{d}\cdot\mathbf{E^h}=d_x\cdot E^h_x$ and $\mathbf{d}\cdot\mathbf{E^v}=d_y\cdot E^v_y$.

From ref.~\onlinecite{ref:Xu_99}, the total power emitted by the dipole in the cavity is $P_{rad}\propto\sum_n{\left|\mathbf{d}\cdot\mathbf{E^n}\right|^2}$, where $\mathbf{d}$ is the dipole moment and $\mathbf{E^n}$ is the (normalized) electric field for mode $n$, evaluated at the dipole location. With the symmetry considerations above, we can write
\begin{equation}
P_{rad}\propto\sum_{n}\left|d_x\cdot E^{h,n}_x\right|^2+\left|d_y\cdot E^{v,n}_y\right|^2=\left|\mathbf{d}\right|^2\left\{\sum_{n}\left|\cdot E^{h,n}_x\right|^2\cos^2\phi+\left|\cdot E^{v,n}_y\right|^2\sin^2\phi\right\},
\label{eq:P_rad}
\end{equation}
where $\phi$  is the dipole orientation - the angle the dipole makes with respect to the x-axis - which we assume to be unknown. Equation~(\ref{eq:P_rad}) thus allows us to determine the dipole emitted power $P_{rad}$ for a dipole positioned anywhere on the $x$ axis, with arbitrary angle given by $\phi$, just based on the $E^h$ and $E^v$ modes which are respectively excited by the $x$ and $y$ dipole components.

As such, we proceed to calculate $P_{rad}$ and the collection efficiency $\eta_{0.4}$ for $NA=0.4$ separately for $x$- and $y$-oriented dipoles located on the $x$-axis at varying distance $x_0$ from the cavity center. We then use eq.(\ref{eq:P_rad}) to determine the range within which the quantities of interest can vary due to the (unknown) azimuthal dipole orientation $\phi$. To verify that this procedure is valid, we also simulate the case $\phi=45^\circ$, and compare the collection efficiency with that obtained through eq.~(\ref{eq:P_rad}) and the $x-$ and $y-$ dipole solutions. As shown in Supplementary Fig.~\ref{fig:dipole45_comp}, the difference between the two types of calculations is  $ \lesssim1~\%$ almost everywhere.

\begin{figure}[h]
\begin{center}
\includegraphics[scale=0.65]{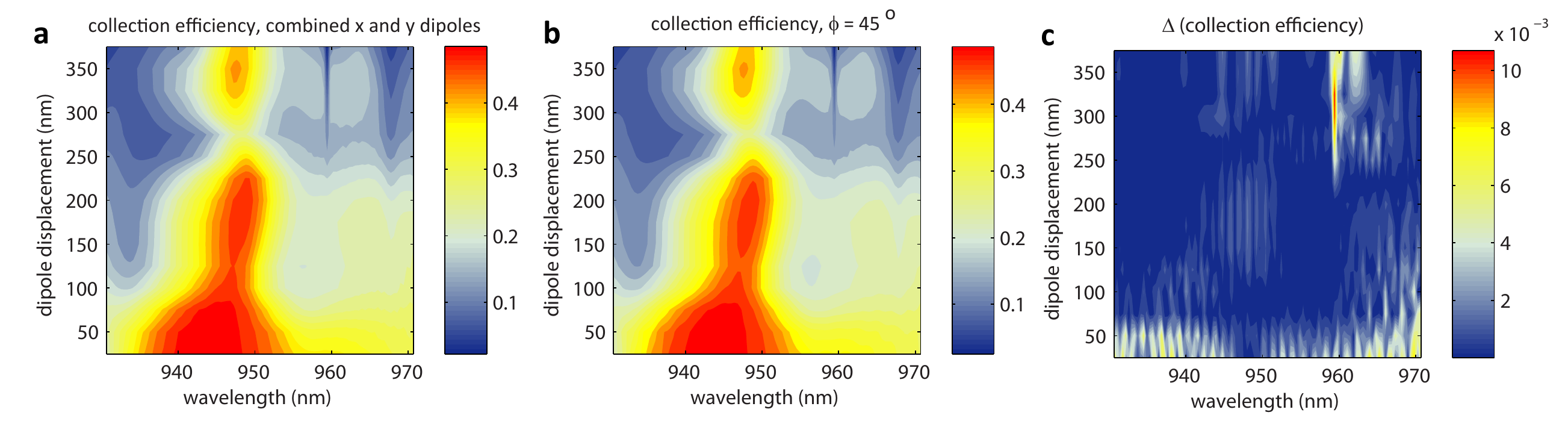}
\caption{Collection efficiency $\eta_{0.4}$ into a $NA=0.4$ objective as a function of wavelength and dipole position along the $x$-axis inside the bullseye cavity. The dipole is on the $xy$-plane and is oriented at an azimuthal angle $\phi=45^\circ$. (a) Results obtained using eq.(\ref{eq:P_rad}) with separate simulations for $x$ and $y$ dipoles individually. (b) Results obtained by simulating a dipole at $\phi=45^\circ$. (c) Absolute value of difference between results shown in panels (a) and (b).}
\label{fig:dipole45_comp}
\end{center}
\end{figure}

In Supplementary Figs.~\ref{fig:Fp}(a) and (b), we show the Purcell Factor $F_p$ as a function of wavelength for $x$- and $y$-dipoles, respectively, located at varying positions along $x$. At a wavelength of $948$~nm, the $x$ dipole couples to the '$h$' resonance shown in Supplementary Fig.~\ref{fig:dipole}(b), while the $y$-dipole couples to the degenerate '$v$' mode in Supplementary Fig.~\ref{fig:dipole}(c), and the Purcell Factor $F_p$ peaks for dipoles at the cavity center. For y-dipoles displaced from the center, $F_{p}$ shows a sequence of satellite peaks observed for increasing distances, which contrasts with the $x$-dipole case.  This can be understood based on the variation of of the $v$ and $h$ field profiles along the $x$-axis (Supplementary Figs.~\ref{fig:dipole}(b)-(c)), as $F_{p}\propto\left|E\right|^2$. A second resonance centered at 957~nm exists that is also excited by dipoles in both orientations, however displays considerably lower Purcell enhancement and collection efficiency (shown later). Supplementary Figs.~\ref{fig:Fp}(c) and (d) show the overall maximum and minimum achievable $F_p$, and the shaded areas in Supplementary Fig.~\ref{fig:Fp}(e) correspond to overall allowed values of $F_p$ as a function of dipole displacement, for three wavelengths around the resonance center. Essentially, these ranges correspond to the uncertainty in our knowledge of $F_p$ due to lack of knowledge of the in-plane dipole orientation. Dotted white lines, on the other hand, correspond to the case $\phi=45^\circ$, which corresponds to an in-plane isotropic dipole.

\begin{figure}[h]
\begin{center}
\includegraphics[scale=0.85]{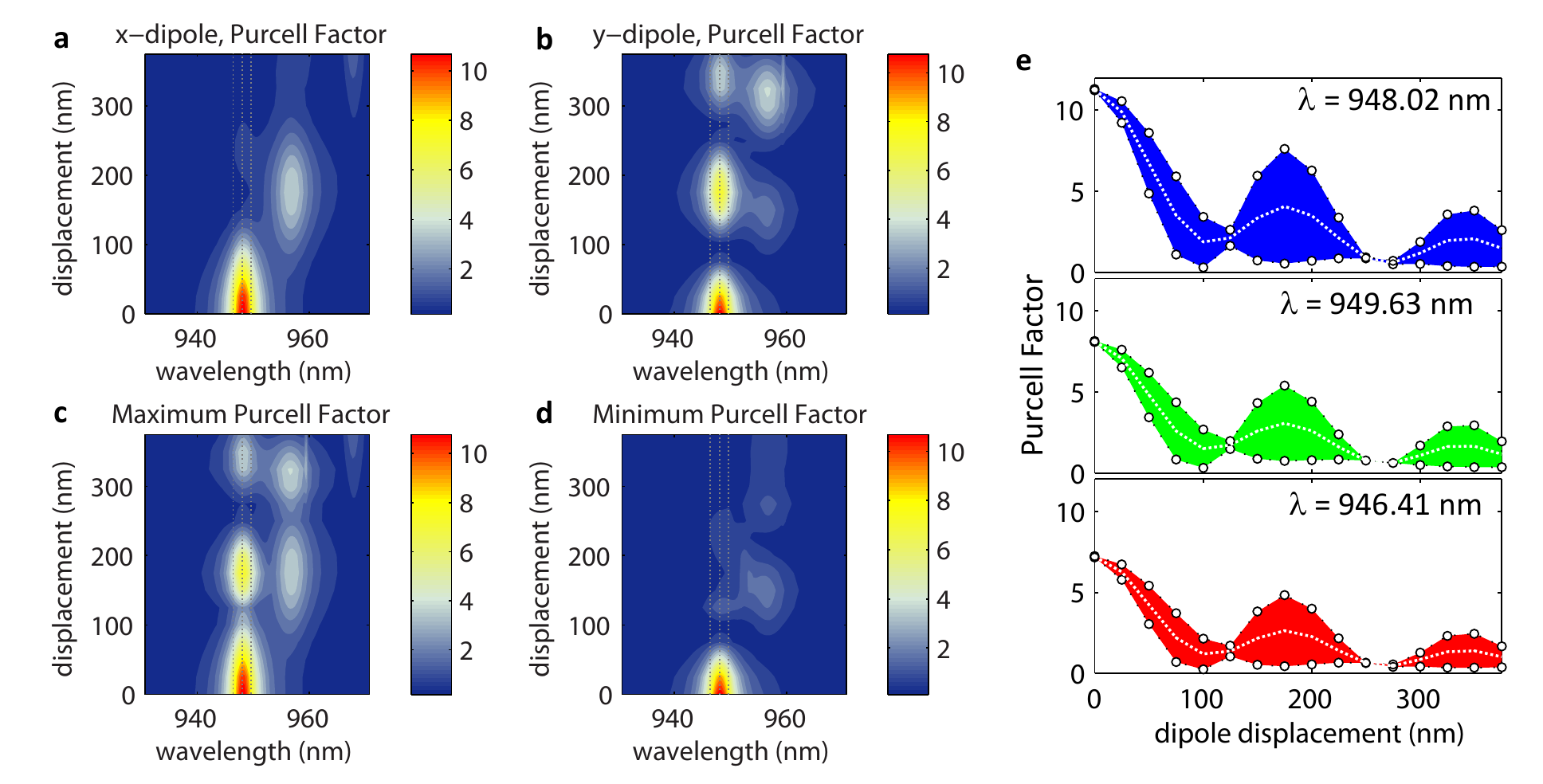}
\caption{Purcell Factor $F_p$ as a function of wavelength and dipole position along the $x$-axis, inside the bullseye cavity. (a) Results for an $x$-oriented dipole; (b) Results for a $y$-oriented dipole; (c) maximum achievable $F_p$; (d) minimum achievable $F_p$; (e) Purcell Factor as a function of dipole displacement from the bullseye cavity center, at resonance ($\lambda=948.02$~nm) and at $\pm$~1.6~nm away [shown as dashed lines in (a)-(d)]. Shaded areas correspond to the uncertainty in $F_p$ due to lack of knowledge of the dipole azimuthal angle $\phi$. The white dotted line corresponds to the case $\phi=45^\circ$.}
\label{fig:Fp}
\end{center}
\end{figure}

\begin{figure}[h]
\begin{center}
\includegraphics[scale=0.65]{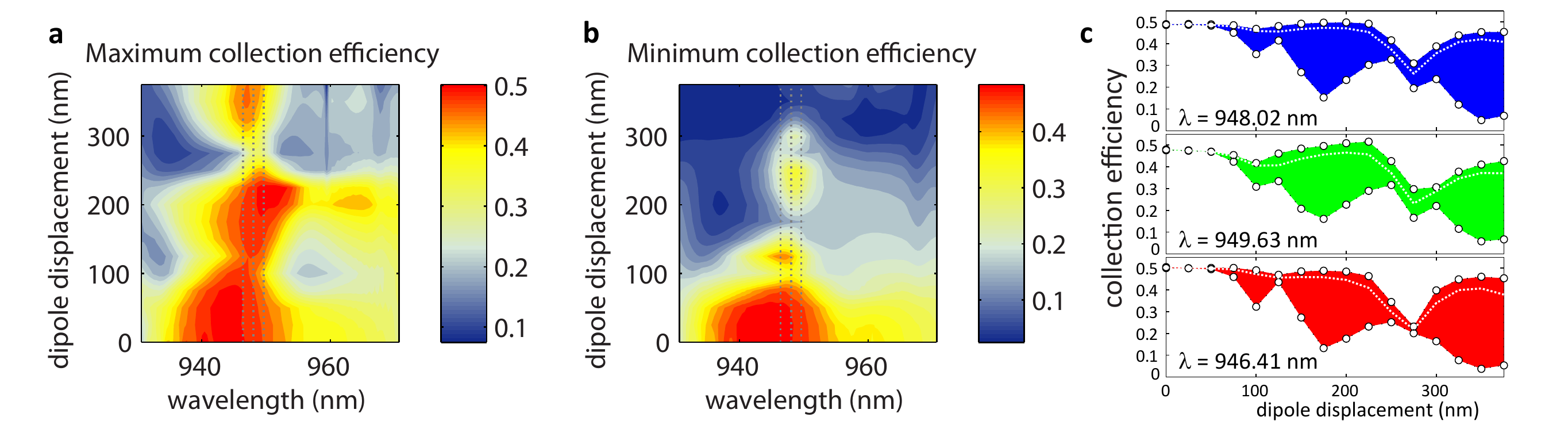}
\caption{Collection efficiency $\eta_{0.4}$ into a $NA=0.4$ optic as a function of wavelength and dipole position along the $x$-axis inside the bullseye cavity. (a) Maximum achievable $\eta_{0.4}$; (b) minimum achievable $\eta_{0.4}$; (c) Collection efficiency as a function of dipole displacement from the bullseye cavity center, at resonance ($\lambda=948.02$~nm) and $\pm$~1.6~nm away [shown as dashed lines in (a) and (b)]. Shaded areas correspond to the uncertainty in $\eta_{0.4}$ due to lack of knowledge of the dipole azimuthal angle $\phi$. The white dotted line corresponds to the case $\phi=45^\circ$.}
\label{fig:eff}
\end{center}
\end{figure}

Supplementary Figs.~\ref{fig:eff}(a) and (b) show the overall maximum and minimum achievable collection efficiency $\eta_{0.4}$, and the shaded areas in Supplementary Fig.~\ref{fig:Fp}(c) correspond to overall allowed values of $\eta_{0.4}$ as a function of dipole displacement, for three wavelengths around the resonance center. White dotted lines are for the $\phi=45^\circ$ case. It is evident that the collection efficiency is a much slower function of both wavelength and dipole displacement than the Purcell factor. As a result, for the QD-cavity wavelength detuning of the device we focus on in the main text (1.6~nm), there is a $\approx~\pm~250$~nm range of dipole positions consistent with the experimentally observed collection efficiency ($48~\%\pm5~\%$) and Purcell Factor ($\approx3$). The lack of knowledge about the QD dipole orientation $\phi$ prevents us from more precisely estimating the location of the emitter within the cavity. For example, if $\phi=45^\circ$, from Supplementary Figs.~\ref{fig:Fp}(e) and Supplementary Fig.~\ref{fig:eff}(c) we can estimate that the dipole is located within 50 nm of the cavity center, in order to display the experimentally observed $F_{p}$ and $\eta_{0.4}$.

We note however that the collection efficiency maximum is shifted with respect to the resonance center by approximately -5~nm, as can be seen in Supplementary Figs.~\ref{fig:Fp}(a)-(d) and Supplementary Figs.~\ref{fig:eff}(a)-(b). This is due to far-field collection efficiency, which is actually asymmetric with respect to the resonance center, being higher by approximately $0.5~\%$ at a maximizing blue-shifted wavelength. While this information is still not sufficient to pinpoint the quantum dot location based on our experimental data, it further corroborates our explanation that the relatively low observed Purcell factors can still exist with high collection efficiencies.

\subsection{Polarization of the light emitted by a dipole embedded within a bullseye cavity}
We now study the polarization properties of the light emitted by a dipole in the bullseye cavity.  In particular, our goal is to understand the degree to which polarization-resolved measurements of the far-field intensity can be used to identify the dipole orientation, which in turn would enable more precise determination of the dipole location from Purcell enhancement and collection efficiency measurements.

The '$h$' and '$v$' bullseye cavity modes overall display major electric field components oriented in the $x$ and $y$ directions, respectively. This can be verified in two ways: plots of $|E_x|^2$ and $|E_y|^2$ for the '$v$' mode in Supplementary Figure~\ref{fig:polar1}(a) show the former to be overall at least an order of magnitude larger than the latter; and the ratio $R_{xy}=\int_{S_{NA}}\mathbf{dS}|E_x|^2/\int_{S_{NA}}\mathbf{dS}|E_y|^2$, where $S_{NA}$ is the  spherical surface corresponding to a $NA=0.4$ cone, is calculated to be $R_{xy}=3.47$. As such, we expect the far-field produced by a dipole at an arbitrary orientation characterized by the azimuthal angle $\phi$ to display some degree of polarization. This degree of polarization can in principle be resolved by introducing of a linear polarizer above the cavity and determining the variation of the transmitted power (collected into a $0.4~NA$ optic) with respect to the polarizer orientation.

\begin{figure}[h]
\begin{center}
\includegraphics[scale=0.65]{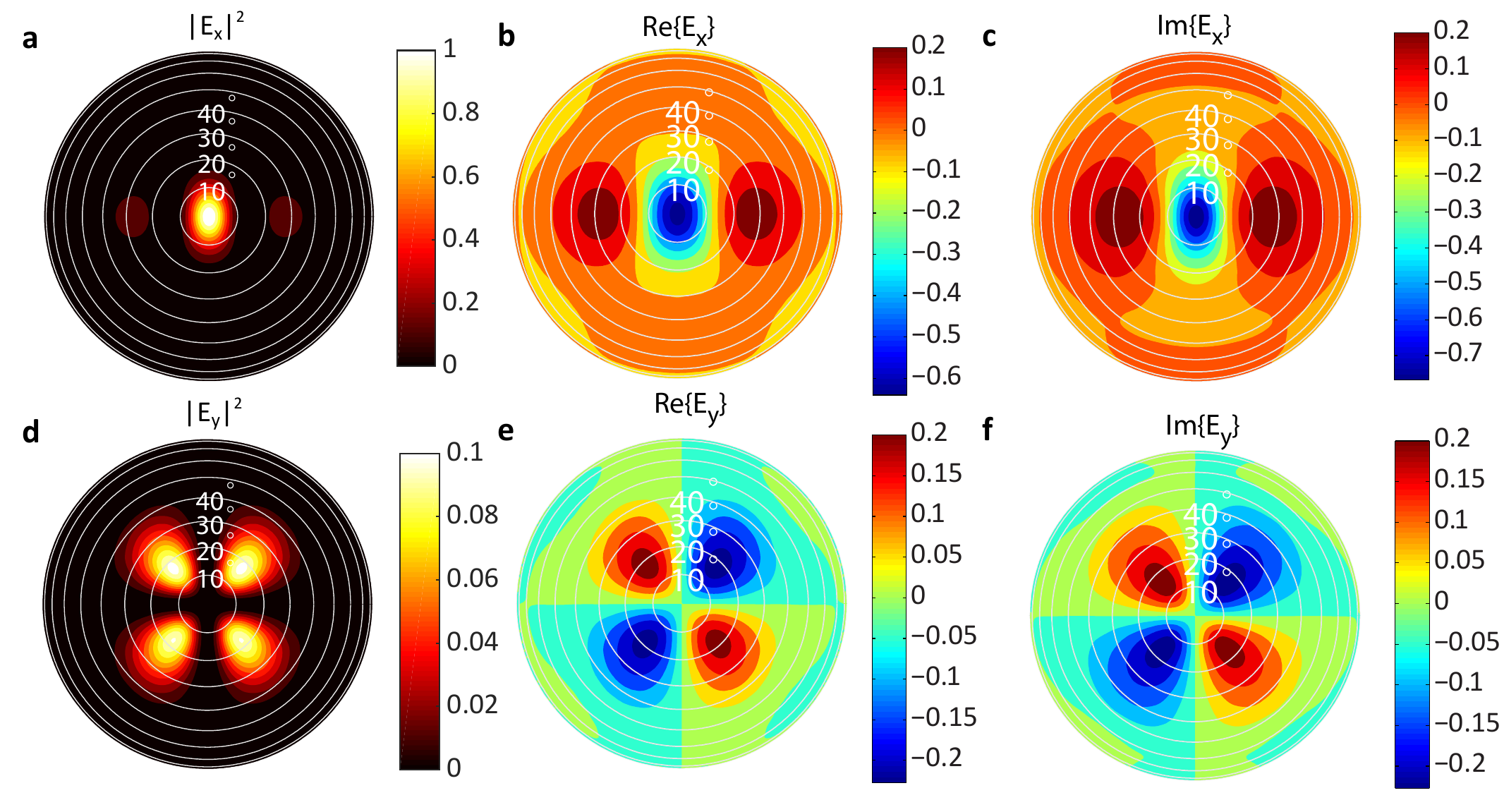}
\caption{(a) Magnitude-square, (b) Real part, and (c) Imaginary part of the $E_x$ far-field for a '$v$' mode. d) Magnitude-square, (e) Real part, and (f) Imaginary part of the $E_y$ far-field for a '$v$' mode. }
\label{fig:polar1}
\end{center}
\end{figure}

To perform this calculation, we first note that the radial component of the far electric field is much smaller than the azimuthal and polar ones ($|\mathbf{E_\rho}|\ll|\mathbf{E_\phi}|,|\mathbf{E_\theta}|$ in spherical coordinates). We then assume that the collection cone is narrow enough that the field at the entrance of the collecting lens can be well represented as $\mathbf{E}=E_x\mathbf{\hat{x}}+E_y\mathbf{\hat{y}}$, where $E_x$ and $E_y$ are the $x$- and $y$-components of the far-field (in other words, we take $E_x$ and $E_y$ to be the transverse components of the far-field). This allows us to use Jones matrix formalism to estimate the power transmitted through the polarizer. We represent a polarizer oriented at an angle $\theta_p$ with respect to the $x$-axis with the Jones matrix
\begin{equation}
\mathbf{M} =  \left[ \begin{array}{cc}
\cos\theta_p & \sin\theta_p  \\
-\sin\theta_p & \cos\theta_p  \end{array}\right]
\left[\begin{array}{cc}
1 & 0  \\
0 & 0  \end{array}\right]
\left[ \begin{array}{cc}
\cos\theta_p & -\sin\theta_p  \\
\sin\theta_p & \cos\theta_p  \end{array} \right] =
\left[ \begin{array}{cc}
\cos^2\theta_p & -\sin\theta_p\cos\theta_p  \\
-\cos\theta_p\sin\theta_p & \sin^2\theta_p  \end{array}\right]
\end{equation}

The transmitted electric field $\mathbf{E_{out}}=\mathbf{ME}$ is, then,
\begin{equation}
\left[ \begin{array}{c}
E_x  \\
E_y  \end{array}\right]_{out}=
\left[\begin{array}{c}
E_x\cos^2\theta_p - E_y\sin\theta_p\cos\theta_p \\
E_y\sin^2\theta_p - E_x\sin\theta_p\cos\theta_p
\end{array}\right].
\end{equation}
The transmitted power is proportional to $|\mathbf{E}|^2=|E_x|^2+|E_y|^2$. If the emitting dipole is at an arbitrary orientation, both '$h$' and '$v$' modes are produced in the cavity, so that, in the far-field, $\mathbf{E}=\alpha_h\mathbf{E^h}+\alpha_v\mathbf{E^v}$ ($\alpha_{h,v}$ represent the dipole coupling strength to the $h$ and $v$ modes). In this case, the resulting expression for the transmitted power consists of a sum of terms $E^i_kE^{j*}_l$, where $i,j\in\{h,v\}$ and $k,l\in\{x,y\}$. In determining the transmitted power, all of these terms are integrated over a portion of a spherical surface which represents the acceptance cone of the collection lens. Because of the cylindrical symmetry of the cavity, the $x$ and $y$ components of the '$h$' and '$v$' fields obey the following symmetry relations (as seen in Supplementary Figs.~\ref{fig:polar1}(b),(c),(d) and (f)): $E^h_x$ and $E^v_y$ are even in $x$ and $y$; $E^v_x$ and $E^h_y$ are odd in $x$ and $y$. Because the integration is performed symmetrically in the $xy$ plane, any cross-term $E^i_kE^{j*}_l$ that results odd in $x$ and $y$ has no contribution to the power; these are cross terms with $\{i\neq j,k=l\}$  and $\{i=j,k\neq l\}$. We can thus write the integrands
\begin{equation}
\left|E_x\right|^2=\cos^4\theta_p\left(\left|E_x^h\right|^2+\left|E_x^v\right|^2\right)+
\cos^2\theta_p\sin^2\theta_p\left(\left|E_y^h\right|^2+\left|E_y^v\right|^2\right)-
2\cos\theta_p\sin^3\theta_p\Re\left\{E_x^hE_y^{v*}+E_x^vE_y^{h*}\right\}
\label{eq:P_pol_Ex2}
\end{equation}
and
\begin{equation}
\left|E_y\right|^2=\sin^4\theta_p\left(\left|E_y^h\right|^2+\left|E_y^v\right|^2\right)+
\cos^2\theta_p\sin^2\theta_p\left(\left|E_x^h\right|^2+\left|E_x^v\right|^2\right)-
2\cos^3\theta_p\sin\theta_p\Re\left\{E_x^{h*}E_y^v+E_x^{v*}E_y^h\right\},
\label{eq:P_pol_Ey2}
\end{equation}
where the substitutions $\mathbf{E^{h,v}\leftarrow\alpha_{h,v}\mathbf{E^{h,v}}}$ were done for simplicity.  We use equations (\ref{eq:P_pol_Ex2}) and (\ref{eq:P_pol_Ey2}) to estimate the collected power that is transmitted through the polarizer, at any polarizer orientation angle . The visibility $V$ can be determined from the maximum and minimum intensities with respect to the polarizer angle as
\begin{equation}
V = \frac{I_{max}-I_{min}}
{I_{max}+I_{min}},
\end{equation}
where $I = \int_{S_{NA=0.4}}\mathbf{dS}\left|\mathbf{E}\right|^2$, where $S_{NA=0.4}$ is the spherical surface corresponding to the $NA=0.4$ collection cone. Once again, the power radiated into $\mathbf{E^h}$ and $\mathbf{E^v}$ by a dipole located at an arbitrary position in the cavity depends on the dipole's orientation; and because the dipole orientation is not known, we can only determine the possible ranges of $V$ at each dipole location. As such, we can only determine the range of achievable visibilities $V$ at each dipole location. This is shown as a function of wavelength in Supplementary Fig.~\ref{fig:V}(s).

These plots indicate the non-monotonic dependence of the visibility on dipole location and orientation.  As a result, a measurement of the visibility, taken together with measurements of the Purcell enhancement and collection efficiency, usually does not provide an unambiguous estimate of the dipole location.

For example, we have measured $V=0.8$ for the device described in detail in the main text.  Based on the quantum dot cavity detuning at the time of this measurement ($-1.6$~nm; lower panel in Supplementary Fig.~\ref{fig:V}(c)), we find that this visibility is consistent with multiple different locations for the dipole, including $~\approx100$~nm away from the cavity center.  While this is consistent with our estimate of dipole location based on the Purcell enhancement and collection efficiency measurements ($<250$~nm from the bullseye center), it does not provide a significantly improved estimate of the dipole location.  Additional measurement techniques (for example, spatially-resolved polarization-dependent far-field measurements) may be required to achieve a better estimate.

\begin{figure}[h]
\begin{center}
\includegraphics[scale=0.9]{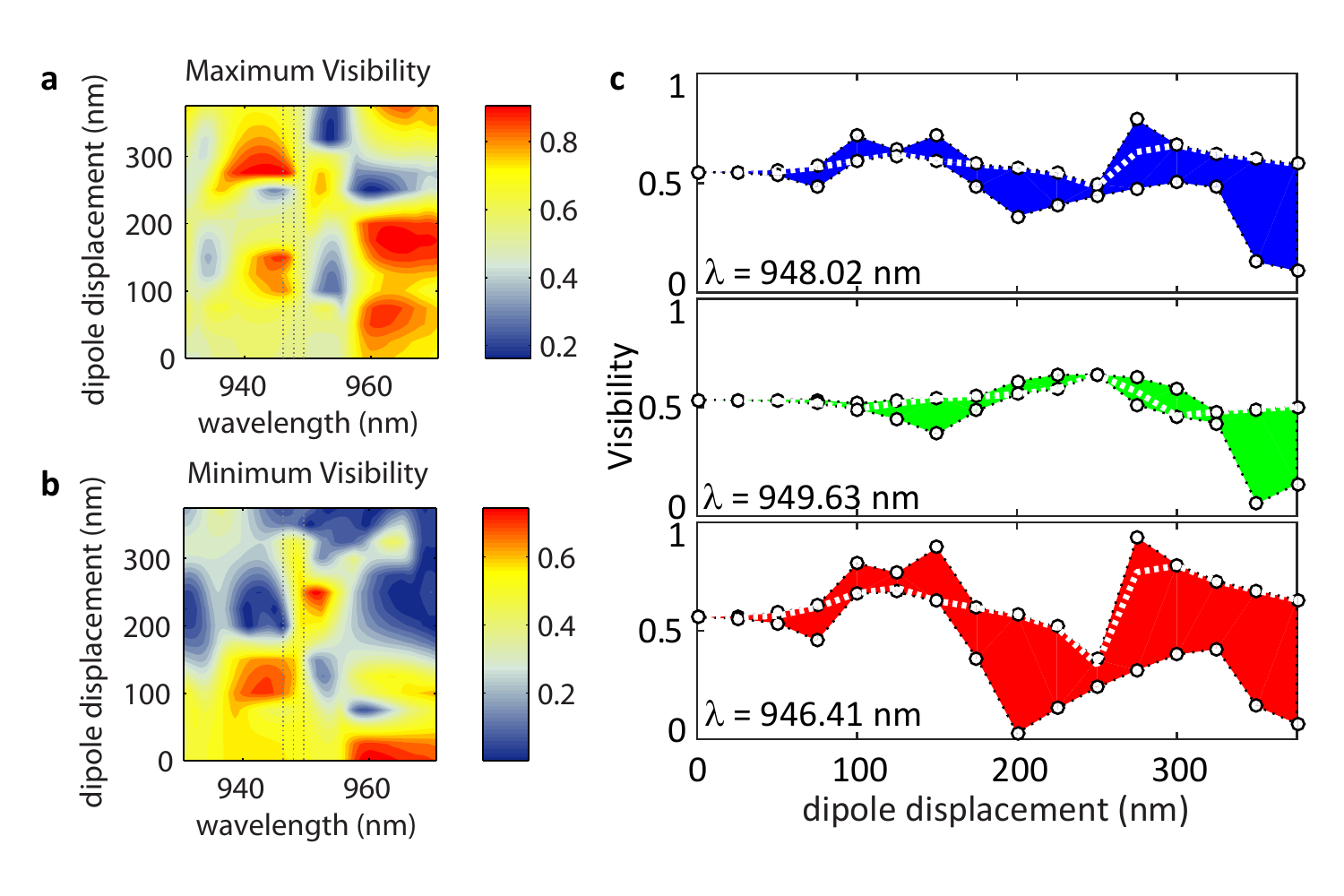}
\caption{Visibility $V$ as a function of wavelength and dipole position along the $x$-axis inside the bullseye cavity. (a) Maximum $V$; (b) minimum $V$; (c) Visibility as a function of dipole displacement from the bullseye cavity center, at resonance ($\lambda=948.02$~nm) and $\pm$~1.6~nm away [shown as dashed lines in (a) and (b)]. Shaded areas correspond to the uncertainty in $V$ due to lack of knowledge of the dipole azimuthal angle $\phi$. The white dotted line corresponds to the case $\phi=45^\circ$.}
\label{fig:V}
\end{center}
\end{figure}


\newpage
\noindent \textbf{\large{Supplementary References}}

\end{document}